\documentclass[preprint,nofootinbib]{revtex4-1}
\usepackage[toc,page]{appendix}
\usepackage{graphicx, amsmath, braket}
\usepackage{mathrsfs, bbm}
\usepackage{caption, subcaption}
\def\lsim{\mathrel{\vcenter{\hbox{$<$}\nointerlineskip\hbox{$\sim$}}}}
\def\gsim{\mathrel{\vcenter{\hbox{$>$}\nointerlineskip\hbox{$\sim$}}}}

\begin{document}

\begin{flushright}
\end{flushright}

\vspace{10mm}
\title{\LARGE Neutrino Sector and Proton Lifetime in a Realistic 
SUSY $SO(10)$ Model}
\author{\bf \large Matthew Severson}
\affiliation{Maryland Center for Fundamental Physics and
Department of Physics, University of Maryland, College Park, MD
20742, USA}
\vspace{10mm}

\begin{abstract}
In this work I present a complete analysis of proton decay in an
$SO(10)$ model previously proposed by Dutta, Mimura, and Mohapatra.
The {\bf 10}, $\overline{\bf{126}}$, and {\bf 120} Yukawa couplings
contributing to fermion masses in this model have well-motivated
restrictions on their textures intended to give favorable results for
proton lifetime as well as a realistic fermion sector without the need
for fine-tuning and for either type-I or type-II dominance in the
neutrino mass matrix. I obtain a valid fit for the entire fermion
sector for both types of seesaw dominance, including $\theta_{13}$ in
good agreement with the most recent data. For the case with type-II
seesaw, I find that using the Yukawa couplings fixed by the successful
fermion sector fit, proton partial lifetime limits are satisfied for
nearly every pertinent decay mode, even for nearly arbitrary values of
the triplet Higgs mixing parameters, with only the $K^+ \bar\nu$ mode
requiring a minor ${\cal O}(10^{-1})$ cancellation in order to satisfy
the experimental limit. I also find a maximum lifetime for that mode
of $\tau(K^+ \bar\nu) \sim 10^{36}$\,years, which should be tested by
eventual experiments. For the type-I seesaw case, I find that all
six pertinent decay modes of interest are satisfied for values of the
triplet mixing parameters giving no major enhancement, with modes
other than $K^+ \bar\nu$ easily satisfied for arbitrary mixing values,
and with a maximum lifetime for $K^+ \bar\nu$ of nearly
$10^{38}$\,years.
\end{abstract}

\maketitle

\section{Introduction} \label{intro} It has been well-established that
a certain class of $SO(10)$ Grand Unified (GUT) models
\cite{minkowski} are capable of elegantly solving some of the most
prominent problems with the Standard Model. One of the more basic yet
intriguing features of these models is the ability to naturally
accommodate a right-handed neutrino, consequently allowing for a
well-motivated implementation of the seesaw mechanism for neutrino
mass \cite{seesaw}, a long-uncontested ansatz that dynamically
explains the smallness of (left-handed) neutrino masses.  The seesaw
was originally implemented in the framework of SUSY $SO(10)$ with only
the {\bf 10}- and {\bf 126}-dimensional Higgs multiplets coupling to
fermions \cite{rabi-aulakh,kuo}; the {\bf 126} vev also plays the
role of breaking $B\!-\!L$ and triggering the seesaw mechanism,
thereby creating a deep mathematical connection between the smallness
of neutrino masses and the other fermion masses. This
seemingly-limited yet elegant approach yielded a realistic neutrino
sector, including an accurate prediction of the value of $\theta_{13}$
\cite{goh-ng,babu-mace}, long before experiments were being done to
measure its properties. This so-called ``minimal'' $SO(10)$ model has
been explored much more thoroughly over the years by many authors with
the arrival of precision measurements
\cite{rabi-babu,babu-mace,10-126a,bajc,goh-ng,10-126b,fukuyama,altarelli},
and it remains a viable predictor of the neutrino sector parameters.

There are however still a number of common difficulties one faces when
attempting to construct a complete and realistic candidate for
unification. Furthermore, these difficulties are continually being
made more severe by new experimental results, which typically manifest
as new lower bounds on the possible existence of some proposed
feature, as part of a disheartening streak of null
results.\footnote{The only recent exception to this null trend was the
discovery \cite{dayabay} of a significantly non-zero (``large'') value
for the reactor mixing angle $\theta_{13}$. Despite the excitement
among experimentalists, the popularity at the time of tri-bimaximal
mixing models \cite{tribim}, which prefer $\theta_{13} \sim 0$, meant
the practical elimination of nearly an entire class of active $SO(10)$
research. See \cite{dms,olddmm,altarelli} for examples of such 
models.\medskip}

Arguably the most problematic feature common to nearly all GUT models
arises when one examines the lifetime of the proton. In all $SO(10)$
models, heavy $SU(5)$-like gauge boson exchanges give rise to
effective higher-dimensional operators that allow for \newpage
quark-lepton mixing and, consequently, nonzero probabilities for
proton decay widths. Furthermore, in SUSY GUT models, several
additional decay modes are available, as each of the GUT-scale Higgs
superfields contains at least one colored Higgs triplet that allows
for proton decay through exchange of Higgsino superpartners.

No one yet knows whether protons do in fact decay at all; so far, the
lower limit on proton lifetime is known to be at least $\sim\!\!
10^{32}$ years, and the partial lifetimes for the various decay modes
have been continually rising through the findings of experiments
\cite{superk}. Thus, if any $SO(10)$ model is to be
trusted, its prediction for the proton lifetime must be at least so
high a number. Most minimal $SU(5)$ models have already been
virtually ruled out by such limits (technically, a few niches of
parameter space do still remain).

There are ways in which the proton lifetime limits can be satisfied
within the framework of a given $SO(10)$ model, but doing so typically
requires substantial fine-tuning to create rather extreme
cancellations ($\gsim \mathcal{O}(10^{-3})$) among the mixing
parameters of the color-triplet Higgsinos exchanged in the decay. The
values of those mixings cannot so far be reasonably recognized as more
than arbitary free parameters, so to expect multiple instances of very
sensitive relationships among them requires putting much faith in
either unknown dynamics or extremely good luck. Restricting the SUSY
vev ratio $v_u/v_d$, conventionally parametrized as $\tan\beta$, to
small values can provide some relief without cancellation for
Higgsino-mediated decay channels, but such an assumption is still ad
hoc and may ultimately be inconsistent with experimental findings;
hence it is strongly preferable to construct a model which is
tractable for any feasible $\tan\beta$.

If however the GUT Yukawas, which are $3\!\times\!3$ matrices in
generation space, have some key elements naturally small or zero, then
extreme cancellations can be largely avoided by eliminating most of
the dominant contributions to proton decay width. A 2013 paper by
Dutta, Mimura, and Mohapatra \cite{dmm1302} proposed such a Yukawa
texture for the $SO(10)$ model including a \textbf{120} coupling in
addition to the \textbf{10} and $\overline{\bf{126}}$ Higgs
contributions to fermion masses. The authors gave a tentative analysis
of mainly leading-order relationships between key fermion fit
parameters and proton partial lifetimes for a neutrino sector arising
from a type-II dominant seesaw mechanism; they also provided a cursory
analysis for a possible type-I solution.

The work I present in this paper verifies the initial analysis of ref.
\cite{dmm1302} by $(a)$ finding a stable numerical fit to all fermion
mass and mixing parameters, including the neutrino sector, where
values are predictions of the model, and $(b)$ then finding adequately
large lifetimes for the dominant modes of proton decay using the
Yukawa couplings fixed by the fermion fit. I grounded the analysis in
conservative assumptions, including large $\tan\beta$, and a
comprehensive calculation relying on as few approximations as
necessary. The pertinent modes of proton decay I checked for
sufficient partial lifetimes are $p \rightarrow K^+ \bar{\nu}$, $K^0
\ell^+$, $\pi^+ \bar{\nu}$, and $\pi^0 \ell^+$, where $\ell = e,\mu$.
I will present solutions for both type-I and II seesaw neutrino
masses.

The results not only give satisfactory predictions for the neutrino
sector based on corresponding charged sector fits, but also adequately
predict sufficiently long-lived protons without relying on the usual
large degree of cancellation. Furthermore, I find that the ansatz is
completely successful in satisfying the proton lifetime limits without
any need for tuning in the type-I seesaw scenario, while a modest
$\mathcal{O}(10^{-1})$ cancellation is still needed in the type-II
case to satisfy the partial lifetime limit of the often-problematic $p
\rightarrow K^+ \bar{\nu}$ mode. This combination of type-I and II
results is precisely contrary to the tentative expectations of the
authors in \cite{dmm1302}; the discrepancy is due mainly to the
unexpected significance of the effect of rotation to mass basis on the
results of the decay width calculations, combined with the numerical
details of the rotation matrices arising from the charged sector mass
and CKM fit. 

The paper is organized as follows: in section \ref{model}, I give an
overview of the $SO(10)$ superpotential and the fermion mass matrices
following from it, followed by the details of the Yukawa texture
ansatz; in section \ref{proton}, I expand further on the model
specifics and examine general GUT proton-decay logistics in order to
derive the needed partial decay widths; in section \ref{fit}, I
present the fermion sector results of the numerical fitting to the
measured masses and mixings; in section \ref{pfit}, I present the
results of the calculation of the important partial lifetimes of the
proton; and in section \ref{conclusion}, I discuss the implications of
the results and give my conclusions.

\section{Details of the Model} \label{model} As mentioned in the
introduction, the $SO(10)$ model in question has \textbf{10}-,
$\overline{\bf{126}}$-, and \textbf{120}-dimensional Yukawa couplings
contributing to fermion masses. The fields are named here as H,
$\overline{\Delta}$, and $\Sigma$, respectively. Thus, the relevant
superpotential terms are
\begin{equation}
  W_Y \ni h_{ij} \Psi_i \Psi_j {\rm H} + f_{ij} \Psi_i \Psi_j
  \overline{\Delta} + g_{ij} \Psi_i \Psi_j \Sigma,
\label{eq:W}
\end{equation}
where $\Psi_i$ is the \textbf{16}-dimensional matter spinor containing
superfields of all the SM fermions (of one generation) plus the
right-handed neutrino, and $i$ is the generation index.

After the GUT symmetry breaking, SM-type $SU(2)$ doublet
representations \linebreak ($\left(\,{\bf 1},{\bf 2},-\frac{1}{2}
\,\right)$ + c.c\,) contained in the decompositions of H,
$\overline{\Delta}$, and $\Sigma$ mix with each other, and also with
the doublets from \textbf{126} (which is needed to preserve SUSY
invariance) and any additional fields present in the model for
GUT-breaking purposes but not contributing to fermion masses, such as
\textbf{210} or \textbf{54}. These doublets come in pairs with
conjugate SM quantum numbers, and each Higgs superfield contains one
or two pairs. The mass matrix $\mathcal{M_D}$ for each set of doublets
is determined by the couplings and vevs of the GUT-scale
superpotential, and so the fields are generally expected to be heavy;
however, one pair must remain light in order to play the role of the
MSSM Higgs doublets $H_{u,d}$. This need requires the imposition of
the fine-tuning condition $Det\,\mathcal{M_D}\sim0$ ({\it i.e.},
$M_{SUSY} \sim 0$ when compared to the GUT scale), which can be
interpreted as the fixing of one parameter in the matrix,
conventionally chosen to be the mass of the \textbf{10}, $M_H$. This
choice will have implications for proton decay analysis that I will
discuss in the next section. In light of this establishment of the
MSSM doublets, the effective Dirac fermion mass matrices can be
written as
\begin{align}
  {\cal M}_u &= \tilde{h}+r_2 \tilde{f}+r_3\tilde{g} \nonumber \\ 
  {\cal M}_d &= \frac{r_1}{\tan\beta}
  (\tilde{h}+\tilde{f}+\tilde{g}) \nonumber \\
  {\cal M}_e &= \frac{r_1}{\tan\beta}
  (\tilde{h}-3\tilde{f}+c_e\tilde{g}) \nonumber \\
  {\cal M}_{\nu_D} &= \tilde{h} - 3 r_2 \tilde{f} + c_\nu \tilde{g},
  \label{eq:mass}
\end{align}
where $1 / \tan \beta$ takes $v_u \rightarrow v_d$ for down-type
fields. Each coupling $\tilde{\lambda}_{ij}$ for $\lambda = h,f,g$ is
related to $\lambda_{ij}$ from eq.\,(\ref{eq:W}) by an absorption of
the SUSY vacuum expectation value (vev) $v_u$ and some function of
elements of the unitary matrices $U^{\cal D}_{I\!J},\, V^{\cal
D}_{I\!J}$ that diagonalize $\mathcal{M_D}$. These mixings are given
in detail with respect to this model in \cite{olddmm} and more
generally in \cite{aulgarg}, but those details are not relevant at
this point in the discussion. The coefficients $r_i$ and $c_\ell$ are
similarly defined as functions of those mixings.

The full neutrino mass matrix is determined by both Majorana mass
terms in the superpotential and the Dirac mass contribution given in
eq.\,(\ref{eq:mass}). The light masses can be generally given by a
combination of the type-I and type-II seesaw mechanisms, involving the
vevs of both left- and right-handed Majorana terms:
\begin{equation}
  {\cal M}_\nu = v_L f - {\cal M}_{\nu_D}\left(v_R f\right)^{-1}\left(
  {\cal M}_{\nu_D}\right)^T,
\label{eq:neu}
\end{equation}
where $v_{L,R}$ are the vevs of the SM-triplet $\overline{\Delta}_{L}$
and singlet $\overline{\Delta}_{R}$ in $\overline{{\bf 126}}$. The
seesaw scale ({\it i.e.}, RH-neutrino scale) coincides with the
$B\!-\!L$ breaking scale and is set by $v_R$; typically $v_L \sim
v_{\rm wk}^2 / v_R$, although it is a free parameter of the model in
principle. We will separately consider cases of type-II ($v_L$ term)
and type-I ($1/v_R$ term) dominance, which can both be readily
accommodated in this model. Note that the presence of the $f$ coupling
in both terms intimately connects the neutrino mass matrix properties
to those of the charged sector matrices, making the model quite
predictive. Also note we will consider only normal mass
hierarchy in this analysis.

The matrices $h$ and $f$ (with tildes or not) are real and symmetric,
and $g$ is pure imaginary and anti-symmetric; hence, the Dirac fermion
Yukawa couplings are Hermitian in general, and their most general
forms can be written as
\smallskip
\begin{gather}
  \tilde{h} = \left(\begin{array}{ccc} 
    h_{11} & h_{12} & h_{13}\\
    h_{12} & h_{22} & h_{23}\\
    h_{13} & h_{23} & M \end{array}\right), \qquad
  \tilde{f} = \left(\begin{array}{ccc}
    f_{11} & f_{12} & f_{13}\\
    f_{12} & f_{22} & f_{23}\\
    f_{13} & f_{23} & f_{33}\end{array}\right), \nonumber \\[4mm]
  \tilde{g} = i \left(\begin{array}{ccc} 
    0 & g_{12} & g_{13} \\
    -g_{12} & 0 & g_{23} \\
    -g_{13} & -g_{23} & 0 \end{array}\right) \label{eq:y0}.
\end{gather}
$M \equiv h_{33} \sim m_t$ is singled out to stress its dominance over
all other elements. The three matrices as written have a total of 15
parameters; taken in combination with $v_L$ as well as the vev and
mixing ratios $r_i$ and $c_\ell$, the model has a total of 21
parameters. Correspondingly, there are in principle 22 measurable
observables, including all masses, mixing angles, and CP violating
phases, associated with the physical fermions, although the three PMNS
phases and one neutrino mass are yet to be observed.  Therefore we
would prefer then to have no more than 18 parameters in the model, and
generally speaking fewer parameters indicates greater predictability.

Furthermore, as I will discuss in more detail shortly, the $d=5$
effective operators that arise in proton decay are $\sim \lambda_{ij}
\lambda_{kl}$ (again $\lambda = h,f,g$); therefore, increasing the
number of $\lambda_{ij}$ that are small or zero will increase the
number of negligible or vanishing contributions to the decay width.
This idea was given thorough consideration in \cite{dmm1302}, and the
couplings suggested by the authors are as follows:
\smallskip
\begin{gather}
  \tilde{h} = \left(\begin{array}{ccc} 
    0 & & \\
    & 0 & \\
    & & M \end{array}\right), \qquad
  \tilde{f} = \left(\begin{array}{ccc}
    \sim 0 & \sim 0 & f_{13}\\
    \sim 0 & f_{22} & f_{23}\\
    f_{13} & f_{23} & f_{33}\end{array}\right), \nonumber \\[4mm]
  \tilde{g} = i \left(\begin{array}{ccc} 
    0 & g_{12} & g_{13} \\
    -g_{12} & 0 & g_{23} \\
    -g_{13} & -g_{23} & 0 \end{array}\right) \label{eq:y}.
\end{gather}
\smallskip Note that $\tilde{h}$ is an explicitly rank-1 matrix, with
$M \sim \mathcal{O}(1)$; thus, at first order, the \textbf{10} Higgs
contributes to the third generation masses and nothing more. This
feature has been explored in models demonstrating a discrete flavor
symmetry in {\it e.g.} \cite{dmmflavor,ddms}, and may therefore be
dynamically motivated.  Taking $f_{12} \sim 0$ is equivalent to a
partial diagonalization of $\tilde{f}$, so it can be done without loss
of generality, and the restriction on $f_{11}$ is clearly
phenomenologically motivated given the small first-generation masses.
As a result of these assumptions, the above Yukawa texture should give
rise to sufficient proton decay lifetimes without the need for the
usual extreme cancellations.

It is further preferred for proton decay that $f_{13},\; g_{12} \ll
1$, although $f_{13}$ plays a role in setting the size of the reactor
neutrino mixing angle $\theta_{13}$, so the above restriction may
create some tension in the fitting.

In carrying out the numerical minimization, I will allow $f_{11}$ and
$f_{12}$ to have small but non-vanishing values,
$\mathcal{O}(10^{-4})$, for the sake of giving accurate
first-generation masses without creating tension in other elements.
The results of that analysis will be discussed in section \ref{fit},
after I discuss the details of calculating proton decay. 

\section{Details of Proton Decay} \label{proton} In addition to the
the SM-doublets present in each of the GUT Higgs superfields, which
contribute to the emergence of $H_{u,d}$ at the SUSY scale, the GUT
fields similarly contain SM-type $SU(3)$ \emph{color-triplets}
($\left(\,{\bf 3},{\bf 1},-\frac{1}{3} \,\right)$ + c.c\,) in their
decompositions. These fields will also mix after the GUT-scale
breaking (again, this mixing includes triplets contained in the Higgs
fields not contributing to fermion masses). Since there is no light
triplet analog to $H_{u,d}$ found in the low-scale particle spectrum,
all of the fields can be heavy, although the decoupling of the
doublet-triplet behavior is a substantial topic itself. The Yukawa
potential in eq.\,(\ref{eq:W}) leads to interactions with these fields
of the forms $h\,{\rm H}_{\overline {\mathcal{T}}}\, (Q L +
U^{\mathcal{C}} D^{\mathcal{C}})$ and~$h\,{\rm H}_{\mathcal{T}}\, (Q Q
+ U^{\mathcal{C}} E^{\mathcal{C}})$, which violate baryon or lepton
number. The fields $\Psi^\mathcal{C} \equiv C \bar{\Psi}^T$ are
left-handed anti-fermion superfields. Note \emph{e.g.} ``$QL$'' is
shorthand for the $SU(2)$-doublet contraction $\epsilon_{\alpha \beta}
Q^\alpha L^\beta$. There are similar interaction terms for
$\Sigma_{\cal T}$ and $\overline\Delta_{\cal T}$; furthermore, two
more exotic types of triplets also lead to $B$- or $L$-violating
vertices, $\left(\,{\bf 3},{\bf 1}, -\frac{4}{3} \right)$ + c.c, which
interact with two up-type or two down-type RH singlet fermions, and
$\left(\,{\bf 3},{\bf 3}, -\frac{1}{3} \right)$ + c.c, with a pair of
LH doublets.

\begin{figure}[t]
\begin{center}
  \includegraphics{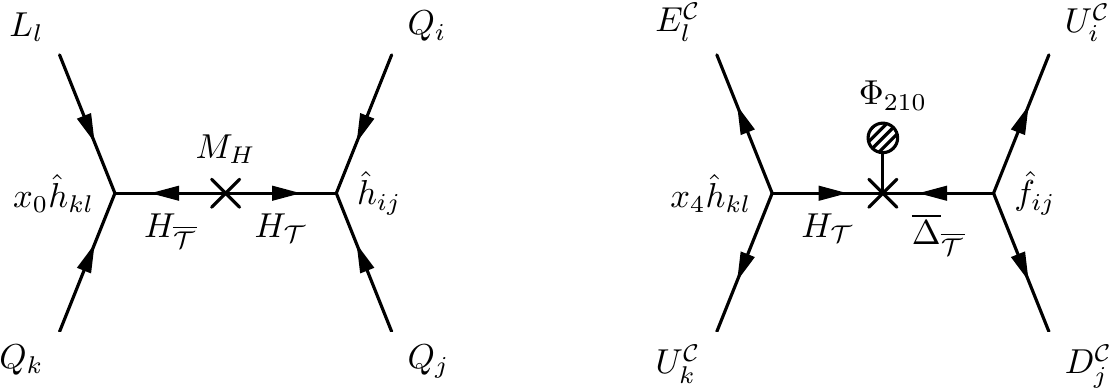}
    \caption{\footnotesize Examples of superfield diagrams that lead to
    proton decay in this model. The hats on the couplings indicate mass
    basis, and the parameters $x_i$ contain the triplet mixing
    information unique to the specific pairing of couplings present in
    each diagram (see below).} 
    \label{fig:susyfd}
\end{center}
\end{figure}

Exchange of conjugate pairs of any these triplets, through a mass term
or interaction with a heavy Higgs field such as {\bf 54} or {\bf 210},
leads to operators that change two quarks into a quark and a lepton;
this is the numerically dominant mechanism through which a proton can
decay into a meson and a lepton; corresponding s-channel decays
through the scalar superpartners of these triplets, as well as
s-channel decays through the $SU(5)$-like gauge bosons $X,Y$ are
suppressed by an additional factor of $1/ M_U$ and so are generally
negligible in comparison.\footnote{The dominant mode in $X$-boson
exchange, $p \rightarrow \pi^0 e^+$, may be comparable if the relevant
threshold corrections are large.} Figure \ref{fig:susyfd} shows
Feynman diagrams for two examples of the operators in question.

\subsection*{The Effective Potential} At energies far below the GUT
scale, the triplet fields are integrated out, giving four-point
effective superfield operators, which give rise in turn to
four-fermion operators. The corresponding $d=5$ superpotential is
\begin{equation}
  {\cal W}_{\Delta B = 1} = \frac{\epsilon_{abc}}{M_\mathcal{T}}
  \left( \widehat{C}^L_{ijkl} Q^a_i Q^b_j Q^c_k L_l +
  \widehat{C}^R_{[ijk]l} U^{\mathcal{C}\,a}_i D^{\mathcal{C}\,b}_j
  U^{\mathcal{C}\,c}_k E^\mathcal{C}_l \right),
\label{eq:effW}
\end{equation}
where $i,j,k,l = 1,2,3$ are the generation indices and $a,b,c = 1,2,3$
are the color indices; the $SU(2)$ doublets are contracted pairwise.
This potential has $\Delta L = 1$ in addition to $\Delta B = 1$ and so
also has $\Delta(B - L) = 0$. $M_\mathcal{T} \sim M_U$ is a generic
GUT-scale mass for the triplets. Note the anti-symmetrization of
$i,k$ in the $C_R$ operator; this is the non-vanishing contribution in
light of the contraction of the color indices. The analogous
anti-symmetry for the $L$ operator is ambiguous in the current
notation, but I will tend to the issue shortly.

The effective operator coefficients $C_{ijkl}$ are of the form
\begin{align}
  C^R_{ijkl} &= x_0 h_{ij} h_{kl} + x_1 f_{ij} f_{kl} +
  x_2 g_{ij} g_{kl} + x_3 h_{ij} f_{kl} +
  x_4 f_{ij} h_{kl} + x_5 f_{ij} g_{kl} \nonumber \\
  & + x_6 g_{ij} f_{kl} + x_7 h_{ij} g_{kl} +
  x_8 g_{ij} h_{kl} + x_9 f_{il} g_{jk} +
  x_{10} g_{il} g_{jk} \nonumber \\
  C^L_{ijkl} &= x_0 h_{ij} h_{kl} + x_1 f_{ij} f_{kl} -
  x_3 h_{ij} f_{kl} - x_4 f_{ij} h_{kl} + y_5 f_{ij} g_{kl} +
  y_7 h_{ij} g_{kl} \nonumber \\
  & + y_9 g_{ik} f_{jl} + y_{10} g_{ik} g_{jl}.
  \label{eq:Cs}
\end{align}
The couplings $h,f,g$ as written correspond to matter fields in the
flavor basis and undergo unitary rotations in the change to mass
basis, as indicated by the hats on $\widehat{C}^{L,R}$ in
eq.\,(\ref{eq:effW}) above; I will save the details of the change of
basis for later in the discussion. The parameters $x_i,y_i \sim
U^{\,\cal T}_{I\!J}, V^{\,\cal T}_{I\!J}$ are elements of the unitary
matrices that diagonalize the triplet mass matrix $\mathcal{M_T}$, or
the corresponding matrices for the exotic triplets. Note that several
identifications have already been made here: $y_{0,1} = x_{0,1}$ and
$y_{3,4} = -x_{3,4}$; the would-be parameters $y_{2,6,8} = 0$. Also
note that $x_0 \sim M_H \sim 1$ is the {\bf 10} mass parameter fixed
by the tuning condition for $M_\mathcal{D}$. The parameters $x_{9,10}$
and $y_{9,10}$ correspond to the exotic triplets; the indices of those
terms are connected in unique ways as a result of the distinct
contractions of fields.

The left-handed term in eq.\,(\ref{eq:effW}) can be further expanded by
multiplying out the doublets as
\begin{equation}
  {\cal W}_{\Delta B = 1} \ni \frac{\epsilon_{abc}}{M_\mathcal{T}}
  \left( \widehat{C}^L_{\{[ij\}k]l} U^a_i D^b_j U^c_k E_l -
  \widehat{C}^L_{\{i[j\}k]l} U^a_i D^b_j D^c_k \mathcal{N}_l \right),
\label{eq:effWL}
\end{equation}
where $\mathcal{N}$ is the left-handed neutrino superfield.  Note that
the coefficients $C^L$ are symmetrized in $i,j$, as a result of the
doublet contractions, and anti-symmetrized in the indices of the
like-flavor quarks, again due to the anti-symmetry of color index
contraction, as discussed above for $C^R$. This anti-symmetry will be
crucial in restricting the number of contributing channels for decay.
Since the symmetrizing of $i,j$ is the same for both types of
left-handed operators, I will suppress its denotation in future
instances to let readability favor the less trivial anti-symmetry.

\begin{figure}[t]
\begin{center}
  \includegraphics{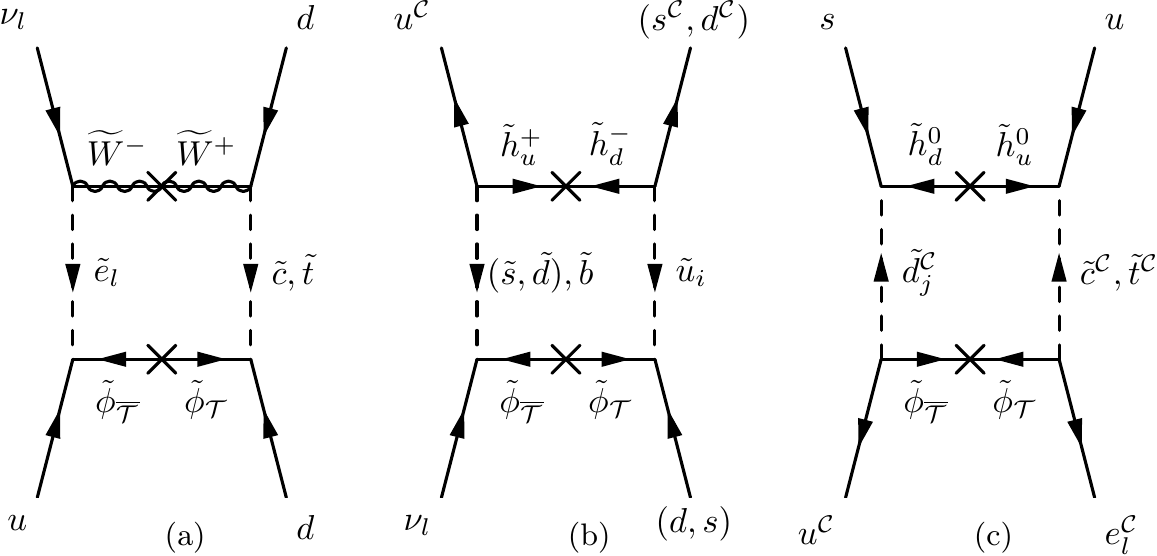}
  \caption{\footnotesize Examples of dressed diagrams leading to proton
  decay in the model. Diagram (a) shows a contribution to $p \rightarrow
  \pi^+ \bar{\nu}_l$; integrating out the triplets gives an effective
  operator of type $C^{L}udue$.  Diagram (b) shows a $C^{L}udd\nu$-type
  operator contributing to $K^+ \bar{\nu}_l$.  Diagram (c) shows a
  $C^{R} u^\mathcal{C} d^\mathcal{C} u^\mathcal{C} e^\mathcal{C}$-type
  operator contributing to $K^0 e_l^+$, for $l=1,2$. Note where more
  than one field is listed, each choice gives a separate contributing
  channel, except for the dependent exchange of $(s \leftrightarrow d)$
  in (b).} 
  \label{fig:compfd}
\end{center}
\end{figure}

\subsection*{Dressing the Operators} As holomorphism of the
superpotential forbids terms like $M_\mathcal{T} \phi_{\mathcal{T}}
\phi_{\overline{\mathcal{T}}}$ for the scalar boson components of the
triplet superfields, diagrams of the type in Figure \ref{fig:susyfd}
can only be realized at leading order through conjugate pairs of {\it
Higgsino} triplet mediators.  Thus, in component notation, each vertex
will be of the form $\lambda\, \tilde{\phi}_\mathcal{T} q\,\tilde{q}$
or similar.  Therefore, the squarks and sleptons must be ``dressed''
with gaugino or (SUSY) Higgsino vertices to give $d=6$ effective
operators of the four-fermion form needed for proton decay. Depending
on the sfermions present, diagrams may in principle be dressed with
gluinos, Winos, Binos, or Higgsinos. Examples of appropriately-dressed
component-field diagrams which give proton decay are shown in Figure
\ref{fig:compfd}.

In the following subsections, I will briefly discuss the implications
for each type of dressing and determine which types will contribute
leading factors in the proton decay width. Note that I will give this
discussion in terms of $\widetilde{B}$, $\widetilde{W}^0$, and
$\tilde{h}_{u,d}^{\pm,0}$, rather than $\widetilde{\chi}^\pm$ and
$\widetilde{\chi}^0_i$, because $(a)$ I will assume a universal mass
spectrum for superpartners to satisfy FCNC constraints, meaning the
mass and flavor eigenstates coincide for the gauge bosons, and $(b)$
the mixing of Higgsinos, while not typically negligible, will result
in chargino or neutralino masses different from $M_{SUSY}$ by
$\mathcal{O}(1)$ factors, as long as gaugino soft masses are relatively
small compared to $M_{SUSY}$; since precise values of such masses are
insofar unknown, and since so many of the SUSY and GUT parameter
values needed for the decay width calculations are similarly unknown,
I will take $m_{\tilde{h}^\pm} \sim m_{\tilde{h}^0} \sim M_{SUSY}
\equiv \mu$ in order to simplify the calculation, especially for
computational purposes.

\subsubsection*{Gluino Dressing} Two limitations are readily apparent
when considering dressing by gluinos. First, the lepton will have to
be a fermion leg in the triplet exchange operator, as in Figure
\ref{fig:compfd} (b) or (c), since a slepton cannot be dressed by a
gluino. Second, since $SU(3)_c$ interactions are
generation-independent, the gluino can only take $\tilde{u}
\rightarrow u$, $\tilde{s} \rightarrow s$, etc. The latter may seem a
fairly innocuous idea on its own, but consider that proton decay to a
kaon or pion will involve operators with one and zero
second-generation quarks as external legs, respectively, with all
others first-generation. Taking these two points together with the
generation-index anti-symmetry of the $C_{ijkl}$ operators, which
implies that $i\neq k$ for the $U_iD_jU_kE_l$ operators and $j\neq k$
for the $U_iD_jD_k\mathcal{N}_l$ operators, one can see by inspecting
a dressed diagram that only diagrams with exactly one each of $U,D,S$
in the triplet operator may be successfully dressed by the gluino.
This constraint implies that gluino dressing can contribute only to $p
\rightarrow K^+ \bar{\nu}$ decay mode; furthermore, the absence of
$UDUE$-type contributions implies no right-handed channels.

\begin{figure}[t]
\begin{center}
  \includegraphics{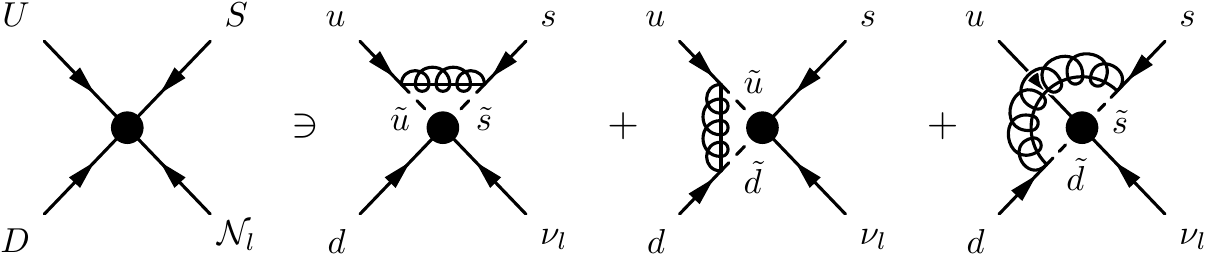}
  \caption{\footnotesize Gluino
  dressings of the $d=5$ operator $M_\mathcal{T}^{-1}
  \widehat{C}^L_{1[12]l} UDS\mathcal{N}$ that would contribute to $p
  \rightarrow K^+ \bar{\nu}_l$; in the limit of universal squark
  masses, the three diagrams sum to zero by a Fierz identity. \\ NOTE:
  gluino mass insertions have been omitted from the diagrams for
  readability.} 
  \label{fig:gludress}
\end{center}
\end{figure}

Taking these constraints into account, and thus looking specifically
at variants of the $UDS\mathcal{N}$ operator, there are three
independent terms we can write \cite{belyaev}, which correspond to
the dressed diagrams shown in Figure
\ref{fig:gludress}:\,\footnote{Each term like ``$(u^a \nu_l)$'' is
actually $(u^a)^T C^{-1} \nu_l$; the details have been suppressed
simply for readability.}
\begin{equation}
  \epsilon_{abc} U^a D^b S^c \mathcal{N}_l \ni \epsilon_{abc}
  \left\{(u^a \nu_l)(\tilde{d}^b \tilde{s}^c) + (d^b
  \nu_l)(\tilde{u}^a \tilde{s}^c) + (s^c \nu_l)(\tilde{u}^a
  \tilde{d}^b)\right\}.
  \label{eq:udsn-q}
\end{equation}
Applying the gluino dressing to each term gives us the following sum
of four-fermion effective operators:
\begin{equation}
  \overset{\tilde{g}}{\longrightarrow} \quad \epsilon_{abc}
  \left(\frac{\alpha_s}{4\pi}\right) \left\{\kappa_1 (u^a \nu_l)(d^b
  s^c) + \kappa_2 (d^b \nu_l)(u^a s^c) + \kappa_3 (s^c \nu_l)(u^a d^b)
  \right\},
  \label{eq:gludr}
\end{equation}
where the parameters $\kappa_a$ contain factors from the
scalar and gluino propagators in the loop integral. The scalar
propagators are different in general; however, recall that I am
assuming universal sfermion mass prescription, meaning that all squark
masses a equal to leading order. In that case, all $\kappa$s are equal
and can be factored out of the brackets. The sum left inside the
brackets is zero by a Fierz identity for fermion contractions
\cite{goh}, and so the contribution from gluino dressing to the $K^+
\bar{\nu}$ decay mode vanishes under the universal mass assumption.

\subsubsection*{Bino Dressing} As with $SU(3)_c$, $U(1)_Y$
interactions are also flavor-diagonal; thus, the same constraints
apply here as in the gluino case, and possible contributions are to
the $K^+ \bar{\nu}$ mode only.

Looking again at the $UDS\mathcal{N}$ operator, for terms in which the
neutrino is a fermion leg, the argument is analogous to that given for
the gluino dressing: the diagrams involved are identical to the three
in Figure \ref{fig:gludress} except with $\tilde{g} \rightarrow
\widetilde{B}$; starting again from expression (\ref{eq:udsn-q})
and applying the Bino dressing, we arrive at an expression similar
to (\ref{eq:gludr}) but containing hypercharge coefficients in
addition to the $\kappa_a$:
\begin{align} \label{eq:binodr-q}
  \overset{\widetilde{B}}{\longrightarrow} \quad \epsilon_{abc}
  \left(\frac{\alpha_1}{4\pi}\right) \{ \kappa_1 Y_d Y_s & (u^a
  \nu_l)(d^b s^c) + \kappa_2 Y_u Y_s (d^b \nu_l)(u^a s^c) \\ & +
  \kappa_3 Y_u Y_d (s^c \nu_l)(u^a d^b)\}; \nonumber
\end{align}
however, $u,d,s \in Q_i$ are all left-handed quarks with $Y =
\frac{1}{6}$, so the hypercharge products factor out, and again the
fermion sum vanishes by the Fierz identity.

\begin{figure}[t]
\begin{center}
  \includegraphics{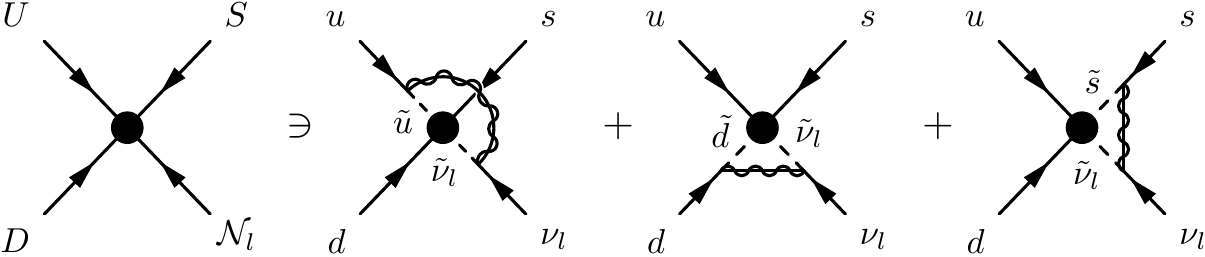}
  \caption{\footnotesize Bino dressings of the $d=5$ operator
  $M_\mathcal{T}^{-1} \widehat{C}^L_{1[12]l} UDS\mathcal{N}$ involving
  a scalar neutrino that would contribute to $p \rightarrow K^+
  \bar{\nu}_l$; again, in the limit of universal squark masses, the
  three diagrams sum to zero by a Fierz identity.  NOTE: Bino mass
  insertions have been omitted from the diagrams for readability.}
  \label{fig:bdress}
\end{center}
\end{figure}

Because leptons carry hypercharge, there are three additional diagrams
one should include in Figure \ref{fig:gludress} if dressing instead by
the Bino, namely, those involving the scalar neutrino; these diagrams
are shown in Figure \ref{fig:bdress}, and the corresponding terms from
the triplet operator are:
\begin{equation}
  \epsilon_{abc} U^a D^b S^c \mathcal{N}_l \ni \epsilon_{abc}
  \left\{(d^b s^c)(\tilde{u}^a \tilde{\nu}_l) + (u^a s^c)
  (\tilde{d}^b \tilde{\nu}_l) + (u^a d^b)(\tilde{s}^c \tilde{\nu}_l)
  \right\}.
  \label{eq:udsn-nu}
\end{equation}
Applying the Bino dressing to each of these terms gives us another sum
of four-fermion effective operators involving hypercharge:
\begin{align} \label{eq:binodr-nu}
  \overset{\widetilde{B}}{\longrightarrow} \quad
  \kappa\,\epsilon_{abc} \left(\frac{\alpha_1}{4\pi}\right) \{Y_u
  Y_\nu & (d^b s^c)(u^a \nu_l) + Y_d Y_\nu (u^a s^c)(d^b \nu_l) \\ & +
  Y_s Y_\nu (u^a d^b)(s^c \nu_l)\}; \nonumber
\end{align}
this group of terms has a different product of hypercharges from that
of (\ref{eq:binodr-q}), but it still has a single common product among
the three terms, so we can again factor it out, leaving us with yet
another vanishing contribution by the Fierz argument.  Hence, the
entire Bino dressing contribution to the $K^+ \bar{\nu}$ mode also
vanishes under the universal mass assumption.  

\subsubsection*{Wino Dressing} As the flavor-diagonal restrictions of
the gluino and Bino also apply to the $\widetilde{W}^0$ but {\it not}
to the $\widetilde{W}^\pm$, they must be considered separately. That
said, one additional restriction applicable in both cases is
the ability to interact with only left-handed particles; thus there
will be no contribution here from the $R$-type operators.

\paragraph*{Neutral Wino.} As noted, dressing with the
$\widetilde{W}^0$ is also restricted to $UDS\mathcal{N}$ contributions
to the $K^+ \bar{\nu}$ mode. The terms to be dressed are the same as
those in the Bino case, given by expressions (\ref{eq:udsn-q}) and
(\ref{eq:udsn-nu}); however, in applying the dressing, we find a kink
in the previous argument:
\begin{align}   
  \overset{\widetilde{W}^0}{\longrightarrow} \quad &
  \kappa\,\epsilon_{abc} \left(\frac{\alpha_2}{4\pi}\right) \{T^3_d
  T^3_s (u^a \nu_l)(d^b s^c) + T^3_u T^3_s (d^b \nu_l)(u^a 
  s^c) + T^3_u T^3_d (s^c \nu_l)(u^a d^b)\} \nonumber \\ 
  =& \,\frac{\kappa\,\epsilon_{abc}}{4}
  \left(\frac{\alpha_2}{4\pi}\right) \{(u^a \nu_l)(d^b s^c) - (d^b
  \nu_l)(u^a s^c) - (s^c \nu_l)(u^a d^b)\},
  \label{eq:winodr-q} \\ \nonumber \\
  \overset{\widetilde{W}^0}{\longrightarrow} \quad &
  \kappa\,\epsilon_{abc} \left(\frac{\alpha_2}{4\pi}\right) \{T^3_u
  T^3_\nu (d^b s^c)(u^a \nu_l) + T^3_d T^3_\nu (u^a s^c)(d^b 
  \nu_l) + T^3_s T^3_\nu (u^a d^b)(s^c \nu_l)\} \nonumber \\ 
  =& \,\frac{\kappa\,\epsilon_{abc}}{4}
  \left(\frac{\alpha_2}{4\pi}\right) \{(d^b s^c)(u^a \nu_l) - (u^a
  s^c)(d^b \nu_l) - (u^a d^b)(s^c \nu_l)\};
  \label{eq:winodr-nu} 
\end{align}
the negative weak isospins carried by the down-type fields prevent us
from using the Fierz identity argument. Thus it seems we have finally
found a non-vanishing contribution to proton decay, albeit to only
this one mode.

There is something yet to be gained from the Fierz identity in this
case: the same zero sum we have seen in the previous cases tells us
that in each expression here, the sum of the two negative terms is
equal to the first term; furthermore, note that the final expressions
in (\ref{eq:winodr-q}) and (\ref{eq:winodr-nu}) are actually
identical. Therefore, we can collect the above contributions into one
expression:
\begin{align}
  \overset{\widetilde{W}^0}{\longrightarrow} \quad 2~\times~ &
  \frac{\kappa\,\epsilon_{abc}}{4} \left(\frac{\alpha_2}{4\pi}\right)
  (-2) \{(u^a s^c)(d^b \nu_l) + (u^a d^b)(s^c \nu_l)\} \nonumber \\ 
  =& ~-\kappa\,\epsilon_{abc} \left(\frac{\alpha_2}{4\pi}\right) 
  \{(u^a s^c)(d^b \nu_l) + (u^a d^b)(s^c \nu_l)\}.
  \label{eq:winodr-sum} 
\end{align}
Including the factors from the triplet operator, we can write an
operator for the entire neutral Wino contribution to $K^+ \bar{\nu}$:
\begin{equation}
  \mathscr{O}_{\widetilde{W}^0} = ~\kappa\,\epsilon_{abc}
  \left(\frac{\alpha_2}{4\pi}\right) M_\mathcal{T}^{-1}
  \widehat{C}^L_{1[12]l}\, \{(u^a s^c)(d^b \nu_l) + (u^a d^b)(s^c
  \nu_l)\},
  \label{eq:wino0-tot} 
\end{equation}
where the sign cancels with that from the $UDD\mathcal{N}$ term in
eq.\,(\ref{eq:effW}). The details of $\kappa$ will be discussed in the
next subsection. Note I could have instead written the above
expression in terms of $(d^b s^c)(u^a \nu_l)$ alone; I chose the
two-operator version because the up-up- and down-down-type pairings in
the latter option are forbidden in Higgsino and charged Wino modes
and so are not seen in the calculation otherwise.

\paragraph*{Charged Wino.} The assumption of universal mass means that
the sfermions are simultaneously flavor and mass eigenstates;
therefore, the would-be unitary rotation matrix for each is simply the
identity, $U^{\tilde{f}} \sim \mathbbm{1}$. As a result, the unitary
matrix present in the fermion-sfermion-Wino couplings is not $V_{\rm
ckm}$, but rather the single unitary matrix corresponding to the
fermion quark rotation. Nonetheless, this rotation allows for the
mixing of generations at the dressing vertices, and the limitations
found on the neutral current dressings are not applicable. This is
quite crucial since it allows for contributions from diagrams with any
squark propagator not forbidden by the anti-symmetry of the
$C^L_{ijkl}$ operator.  Proton decay modes involving neutral Kaons or
pions, which have $u\bar{u}$ or $d\bar{d}$ as external quarks, would
be intractable without generation mixing. Such mixing will of course
come at the expense of suppression from an off-diagonal element in the
pertinent unitary matrix, which will typically be
$\mathcal{O}(10^{-2\mbox{-}3})$; hence, one can begin to see an
indication of why the $K^+ \bar{\nu}$ mode is so dominant in the full
proton decay width.

One additional constraint on charged Wino dressing involves the Wino
mass insertion. Unlike the gauginos discussed so far, $W^\pm$ are the
antiparticles of \emph{each other}, rather than either being its own
antiparticle. As a result, the Wino mass term is of the form
$M_{\widetilde{W}} \widetilde{W}^+ \widetilde{W}^-$; in order to
involve one $\widetilde{W}^+$ and one $\widetilde{W}^-$ in the
dressing, the two sfermions involved must be of opposite $SU(2)$
flavor. As a result, triplet operators of the form $u \tilde{d} u
\tilde{e}$, $\tilde{u} d \tilde{u} e$ (or the RH equivalents), $u
\tilde{d} \tilde{d} \nu$, and $\tilde{u} d d \tilde{\nu}$ do not
contribute.

Beyond these constraints, the generational freedom of the sfermions
leads to numerous contributions to each of the crucial decay modes,
$K^+ \bar{\nu}$, $K^0 \ell^+$, $\pi^+ \bar{\nu}$, and $\pi^0 \ell^+$,
where $\ell = e,\mu$. In particular the $UDUE$- and
$UDD\mathcal{N}$-type operators each contribute to {\it each} mode
through multiple channels. A list of all such contributions would
likely be overwhelming to the reader no matter how excellent my
choices of notation, but one can find the relevant diagrams in
Appendix \ref{fds}.

\subsubsection*{Higgsino Dressing}

When compared to the others, Higgsino dressing is wildly
unconstrained. First, the low-scale Yukawa couplings governing the
fermion-sfermion-Higgsino interactions couple a left-handed field to a
right-handed one, so clearly the dressing can be applied to both
$C^L$- and $C^R$-type triplet operators. Also, since charged and
neutral Higgsinos couple through the same Yukawas, both types of
interactions can mix generations, meaning the generation-diagonal
constraints on the rest of the neutral-current dressings do not apply
to $\tilde{h}^0_{u,d}$. The only previously-mentioned restriction that
{\it does} apply is, like the charged Wino, the mass term for the SUSY
Higgs couples $H_u$ to $H_d$, so it therefore cannot contribute
through the triplet operators with sfermions of like $SU(2)$ flavor.
One remaining minor restriction is that we will not see the triplet
operator $\tilde{u}du\tilde{e}$ dressed by $\tilde{h}^\pm$ nor
$u\tilde{d}d\tilde{\nu}$ dressed by $\tilde{h}^0$ because each would
result in an outgoing left-handed anti-neutrino.

One can find cases in the literature ({\it e.g.} \cite{goh}) of
Higgsino-dressed contributions being counted as negligible when
compared to those from the Wino; this is usually because if one
exchanges the $g_2^2\,V_{\mathrm{Cabibbo}}$ found in a typical
dominant Wino contribution for a $y^u_{ii'}\, y^d_{kk'}\, \tan\beta$
found in a typical dominant Higgsino contribution, the resulting value
will be smaller by at least a factor of $\mathcal{O}(10)$. Of course
one makes several assumptions in such a comparison: $\mu \sim
M_{\widetilde{W}}$ for one, but additionally that $(a)$ $\tan\beta$ is
small or moderate, and $(b)$ the $C_{ijkl}$ coefficients are usually of
roughly the same magnitude for any combination of $i,j,k,l$ present.

For this analysis, though, neither assumption is valid: I have already
mentioned that I will consider large $\tan\beta$ for maximal
applicability; furthermore, due to the rank-1 texture of the $h$
coupling and the related sparse or hierarchical textures of $f$ and
$g$ as shown in eq.\,(\ref{eq:y}), many of the $C_{ijkl}$ are small or
zero, creating large disparities between the values from one
contribution to the next. This discrepancy from expectation is further
enhanced by the tendency for the unitary matrices $U^f$, which
give the off-diagonal suppressions at the dressing vertices in this
model, to individually deviate from the hierarchical structure of
$V_{\rm ckm}$.

To see the extent to which these two properties can lead to surprises
in numerical dominance, consider that, for example, I find $C^L_{1213}
\sim C^L_{3213}\,U^d_{31}$; one might expect that $U^d_{31} \sim
V_{ub}$ and therefore the former term is much larger than the
latter, but in fact neither assumption is accurate. 

As a result of these model characteristics, I find that the dominant
contributions from Higgsino-dressed diagrams are generally comparable
to those from Wino-dressed diagrams. This statement further applies to
contributions from {\it right-handed} operators as well. Thus I made
no {\it a priori} assumptions about which of the $C^L$- or $C^R$-type
Higgsino-dressed contributions might be excluded as negligible.

Because both the $U^\mathcal{C} D^\mathcal{C} U^\mathcal{C}
E^\mathcal{C}$ operators and the $\tilde{h}^0_{u,d}$ dressing
contribute to all of the pertinent decay modes, the complete list of
channels dressed by the Higgsino is considerably more plentiful than
that of the Wino and would again, I suspect, be of no more than
marginal use to any but the most involved reader. Again though one
can find all of the pertinent diagrams in Appendix \ref{fds}.

\subsection*{Building the Partial Decay Width Formulae} As I discussed
above in the Higgsino dressing subsection, the Yukawa texture seen in
eq.\,(\ref{eq:y}) leads to $(a)$ unusually extreme variation in the
sizes of the $C_{ijkl}$ coefficients, depending strongly on the index
values present, and $(b)$ textures for the unitary matrices $U^f$
which deviate substantially from that of $V_{\rm ckm}$. The
repercussions of these features clearly extend beyond affecting the
relative size of Wino and Higgsino channel contributions. For one, the
off-diagonal suppressions $U^f_{kk'}$ present in most charged Wino
diagrams can not be dependably approximated as $V^{\rm ckm}_{kk'}$;
fortunately, the GUT-scale $U^f$ are fixed by the fermion fitting, and
since the running of such unitary matrices is small, I can simply use
them at the $\widetilde{W}^\pm$ vertices as reasonable approximations
to their low-scale counterparts.

Another complication due the Yukawa texture is the disturbance of
typically useful assumptions about which channels dominate the
calculation. Such assumptions include dominance of Higgsino channels with
$\tilde{t},\tilde{b},\tilde{\tau}$ intermediate states or Wino
channels $\propto V_{ii}$ or $V_{\mathrm{Cabibbo}}$. In the absence of
the validity of any such simplification, I am compelled to presume
that {\it any} channel might be a non-negligible contribution to decay
width.

Thus, I initially treated all possible channels as potentially
significant; however, in the interest of saving considerable
computational time, I chose an abridged set of contributions to
include in my numerical analysis through inspection of tentative
calculations, although my threshold for inclusion was quite
conservative. It seemed to me that conventional methods of keeping
only the most dominant terms for calculation might easily lead to
drastically underestimated decay widths, in that if I exclude ten
``negligible'' terms smaller than leading contributions by a factor of
ten, then I have evidently excluded the equivalent of a leading
contribution. To fully avoid such folly, I used a cutoff of roughly
1/50 for exclusion, and made cuts on a per-triplet-operator basis,
which translates to three or four significant figures of precision in
the decay widths.

The Feynman diagrams for all non-vanishing channels of proton decay
for the $K^+ \bar{\nu}_l$, $K^0 \ell^+$, $\pi^+ \bar{\nu}_l$, and
$\pi^0 \ell^+$ modes are catalogued in Appendix \ref{fds}.

Calculation of a proton partial decay width can be broken into three
distinct parts. The first part is the evaluation of the ``internal'',
$d=6$ dressed diagrams discussed in the previous subsection; each
diagram corresponds to an effective operator of the form $X\,qqq\ell$,
where $X \sim M^{-1}_\mathcal{T}\, C_{ijkl}\,$\dots~is a numerical
coefficient unique to each decay channel. Note that here each $q$ is a
single quark fermion, not a doublet. The second part is the evaluation
of a hadronic factor that quantifies the conversion of the three
external quarks of a dressed diagram--plus one spectator quark--into a
proton and a meson. The third and final part is the evaluation of the
``external'' effective diagram for $p \rightarrow \mathrm{M} \bar\ell$
giving the decay width of the proton . I will go through the details
of each stage before giving the resulting decay width expressions.

\paragraph*{Evaluating the Dressed Operators.} The evaluation of one
such dressed $d=6$ box diagram involves calculating the loop integral
but no kinematics, because the physical particles carrying real
momenta here are the proton and the meson, not the quarks. The loop
factor is not divergent and is of the same general form for every
channel; furthermore, as the heavy triplets are common to all diagrams
and the sfermion masses are assumed to be equal, the only factors in
the loop that vary from one channel to the next are the couplings and
masses associated with either the Wino or Higgsino. The remaining
variation from one diagram to the next depends entirely on the
particle flavors, which is apparent in the external fermions and
encoded in the $C_{ijkl}$ coefficients and the unitary matrices
involved in rotation to mass basis. Thus, I can write the operator for
any pertinent diagram as a generic Wino- or Higgsino coefficient times
one of several flavor-specific sub-operators; the forms of the general
operators are

\begin{equation}
  \mathscr{O}_{\widetilde{W}} = \left(\frac{i \alpha_2}{4\pi}\right)
  \left(\frac{1}{M_\mathcal{T}}\right)\, I\left(M_{\widetilde{W}},
  m_{\tilde{q}}\right) \mathscr{C}_{\widetilde{W}}^{\mathcal{A}}
\end{equation}
and
\begin{equation}
  \mathscr{O}_{\tilde{h}} = \left(\frac{i}{16\pi^2}\right)
  \left(\frac{1}{M_\mathcal{T}}\right)\, I\left(\mu,
  m_{\tilde{q}}\right) \mathscr{C}_{\tilde{h}}^{\mathcal{A}},
\end{equation}
where\footnote{One might notice that this expression for $I(a,b)$
differs from what is usually given in the literature for analogous
proton decay expressions; the discrepancy is due to my inclusion of
the universal mass assumption prior to evaluating the loop integral.
\vspace{2mm}}
\begin{equation}
  I(a,b) = \frac{a}{b^2\!-\!a^2} \left\{\,1\, +\,
  \frac{a^2}{b^2\!-\!a^2} \log\left(\frac{a}{b}\right) \right\},
  \nonumber
\end{equation}
and the sub-operators $\mathscr{C}^{\mathcal{A}}$ are\footnote{I do
not list the neutral Wino operator again here, but looking back at
eq.\,(\ref{eq:wino0-tot}), we can see that $\kappa = I
\left(M_{\widetilde{W}}, m_{\tilde{q}}\right)$.}
\begin{align}
  \mathscr{C}_{\widetilde{W}}^I &= \frac{1}{2} (u^T\, C^{-1}\, d_j)\,
  \widehat{C}^L_{[ij1]l}\, U^d_{ii'}\, U^\nu_{ll'}\, (d^T_{i'}\,
  C^{-1}\, \nu_{l'}) \nonumber \\ \mathscr{C}_{\widetilde{W}}^{I\!I}
  &= \frac{1}{2} (u^T\, C^{-1}\, e_l)\, \widehat{C}^L_{[1jk]l}\,
  U^d_{kk'}\, U^u_{j1}\, (d^T_{k'}\, C^{-1}\, u) \nonumber \\
  \mathscr{C}_{\widetilde{W}}^{I\!I\!I} &= -\frac{1}{2} (u^T\,
  C^{-1}\, d_{k})\, \widehat{C}^L_{1[jk]l}\, U^u_{j1}\, U^e_{ll'}\,
  (u^T\, C^{-1}\, e_{l'}) \nonumber \\
  \mathscr{C}_{\widetilde{W}}^{I\!V} &= -\frac{1}{2} (d_j^T\, C^{-1}\,
  \nu_l)\, \widehat{C}^L_{i[jk]l}\, U^d_{ii'}\, U^u_{k1}\, (d^T_{i'}\,
  C^{-1}\, u)
  \label{eq:CWops}
\end{align}
for the (charged) Wino,
\begin{align}
  \mathscr{C}_{\tilde{h}^\pm}^{I} &= (u^T\, C^{-1}\, e_l)\,
  \widehat{C}^L_{[1jk]l}\: y^{d\, \dagger}_{kk'}\, y^{u\,
  \dagger}_{j1}\, (d^{\,\mathcal{C}\,T}_{k'}\, C^{-1}\,
  u^{\mathcal{C}}) \nonumber \\ 
  \mathscr{C}_{\tilde{h}^\pm}^{I\!I} &= -(u^T\, C^{-1}\, d_{k})\,
  \widehat{C}^L_{1[jk]l}\: y^{u\, \dagger}_{j1}\, y^{e\,
  \dagger}_{ll'}\, (u^{\mathcal{C}\,T}\, C^{-1}\,
  e^{\mathcal{C}}_{l'}) \nonumber \\
  \mathscr{C}_{\tilde{h}^\pm}^{I\!I\!I} &=
  -(d_j^T\, C^{-1}\, \nu_l)\, \widehat{C}^L_{i[jk]l}\: y^{d\,
  \dagger}_{ii'}\, y^{u\, \dagger}_{k1}\,
  (d^{\,\mathcal{C}\,T}_{i'}\, C^{-1}\, u^{\mathcal{C}}) \nonumber \\ 
  \mathscr{C}_{\tilde{h}^\pm}^{I\!V} &= (u^{\mathcal{C}\,T}\, C^{-1}\, 
  d^{\,\mathcal{C}}_j)\, \widehat{C}^R_{[ij1]l}\: y^u_{ii'}\, 
  y^e_{ll'}\, (d^T_{i'}\, C^{-1}\, \nu_{l'}) \nonumber \\
  \mathscr{C}_{\tilde{h}^\pm}^V &= (u^{\mathcal{C}\,T}\, C^{-1}\,
  e^{\mathcal{C}}_l)\, \widehat{C}^R_{[1jk]l}\: y^u_{kk'}\, y^d_{j1}\,
  (d^T_{k'}\, C^{-1}\, u)
  \label{eq:Chpops}
\end{align}
for the charged Higgsino, and
\begin{align}
  \mathscr{C}_{\tilde{h}^0}^{I} &= -(u^T\, C^{-1}\, d_{k})\,
  \widehat{C}^L_{[ij1]l}\: y^{u\, \dagger}_{i1}\, y^{e\,
  \dagger}_{ll'}\, (u^{\mathcal{C}\,T}\, C^{-1}\,
  e^{\mathcal{C}}_{l'}) \nonumber \\
  \mathscr{C}_{\tilde{h}^0}^{I\!I} &= -(u^T\, C^{-1}\, e_l)\,
  \widehat{C}^L_{[1jk]l}\: y^{d\, \dagger}_{kk'}\, y^{u\,
  \dagger}_{j1}\, (d^{\,\mathcal{C}\,T}_{k'}\, C^{-1}\,
  u^{\mathcal{C}}) \nonumber \\ 
  \mathscr{C}_{\tilde{h}^0}^{I\!I\!I} &=
  (d_j^T\, C^{-1}\, \nu_l)\, \widehat{C}^L_{i[jk]l}\: y^{u\, \dagger}_{i1}\,
  y^{d\, \dagger}_{kk'}\, (u^{\mathcal{C}\,T}\, C^{-1}\, 
  d^{\,\mathcal{C}}_{k'}) \nonumber \\ 
  \mathscr{C}_{\tilde{h}^0}^{I\!V} &= -(u^{\mathcal{C}\,T}\, C^{-1}\, 
  d^{\,\mathcal{C}}_j)\, \widehat{C}^R_{[ij1]l}\: y^u_{i1}\, y^e_{ll'}\,
  (u^T\, C^{-1}\, e_{l}) \nonumber \\
  \mathscr{C}_{\tilde{h}^0}^V &= -(u^{\mathcal{C}\,T}\, C^{-1}\,
  e^{\mathcal{C}}_l)\, \widehat{C}^R_{[1jk]l}\: y^u_{k1}\, y^d_{jj'}\,
  (u^T\, C^{-1}\, d_{j'}) 
  \label{eq:Ch0ops}
\end{align} 
\noindent for the neutral Higgsino, where I have suppressed the color
indices everywhere. Again the hats on $\widehat{C}^{L,R}$ indicate
$\hat{h},\hat{f},\hat{g}$ are rotated to the mass basis, which I will
discuss in detail shortly. Note that $UDUE$ and $UDD\mathcal{N}$
operators generally differ by a sign, as do diagrams dressed by
$\tilde{h}^\pm_{u,d}$ and $\tilde{h}^0_{u,d}$; the latter difference
arises from the $SU(2)$ contraction in the SUSY Higgs mass term. These
sign differences create the potential for natural cancellation within
the absolute squared sums of interfering diagrams, and even for
cancellation of entire diagrams with each other in some cases. Also
note that the Yukawa couplings are Hermitian in this model, hence the
distinction above between $y^f$ and $y^{f\,\dagger}$ is not relevant
for this work.

I utilized two additional observations to simplify the implementation
of the above operators. First, I took values for the superpartner
masses such that $\mu,M_{\widetilde{W}} \ll m_{\tilde{q}}$, which
imples $I(a,b) \simeq a/b^2$. Also, because I'm only interested in the
combined contribution of the three neutrinos, and because the total
contribution is the same whether one sums over flavor states or mass
states, I made the replacement $U^\nu_{ll'} \rightarrow \delta_{ll'}$
for $\mathscr{C}_{\widetilde{W}}^I$ and took $l = l'~ \Rightarrow~
y^e_{ll'} = m^e_l/v_d$ for $\mathscr{C}_{\tilde{h}^\pm}^{I\!V}$.

Since the unitary matrices $U^f$ do not appear in the SM (+ neutrino
sector) Lagrangian except in the CKM and PMNS combinations, the
non-diagonal SUSY Yukawas $y^f$ present in the
$\mathscr{C}^{\mathcal{A}}$ are not physically determined.
Fortunately in our GUT model full {\it high}-scale Yukawas are defined
by the completely determined fermion sector. Furthermore, it is known
that unitary matrices such as the CKM matrix experience only very
slight effects due to SUSY renormalization. Thus, since the low-scale
masses are of course known, I can define good approximations to the
SUSY Yukawas needed by using the high-scale $U^f$ to rotate
the diagonal mass couplings at the proton scale, divided by
the appropriate vevs:
\begin{equation}
  y^{u} = \frac{1}{v_u}\: U_u\, \left(\mathcal{M}^{\rm wk}_u\right)^D\,
  U_u^\dagger, \nonumber
\end{equation}
where $v_u = v_{\rm wk} \sin \beta$, or, in component notation,
\begin{equation}
  y^{u}_{ij} = \frac{1}{v_u}\: \sum\limits_k\, m^u_k\, U^u_{ik}\,
  U^{u\,*}_{jk}.
\end{equation}
I can similarly write
\begin{align}
  y^{d}_{ij} = \frac{1}{v_d}\: \sum\limits_k\, m^d_k\, U^d_{ik}\,
  U^{d\,*}_{jk} \nonumber \\
  y^{e}_{ij} = \frac{1}{v_d}\: \sum\limits_k\, m^e_k\, U^e_{ik}\,
  U^{e\,*}_{jk}, \nonumber
\end{align}
where $v_d = v_{\rm wk} \cos \beta$. Mass values used were taken from
the current PDG \cite{pdg}; light masses are run to the 1-GeV scale,
top and bottom masses are taken on-shell. Note that since the Yukuwa
factors always appear in pairs of opposite flavor in the Higgsino
operators, and since $\frac{1}{\sin\beta\cos\beta} \simeq \tan\beta$
for large $\beta$, the Higgsino contributions to proton decay $\sim
\frac{\tan^2\beta}{v_{\rm wk}^4}$ for this model.

There are generally two distinct mass-basis rotations possible for
each of the $UDUE\,\mbox{-}$, $UDD\mathcal{N}$-, and $U^{\mathcal{C}}
D^{\mathcal{C}} U^{\mathcal{C}} E^{\mathcal{C}}$-type triplet
operators; the difference between the two depends on whether the
operator is ``oriented'' ({\it i.e.}, in the diagram) such that the
lepton is a scalar. For a given orientation, a unitary matrix
corresponding to the fermionic field at one vertex in the triplet
operator will rotate every coupling present in ${C}^{L,R}$ pertaining
to that vertex; an analogous rotation will happen for the other vertex
in the operator. For example, looking at the $\pi^+ \bar{\nu_l}$
channel in Figure \ref{fig:compfd}(a), every coupling $\lambda_{ij}$
from $C^L_{ijkl}$ present at the $\tilde{\phi}_{\mathcal{T}}$ vertex
will be rotated by some form of $U^d$; similarly all $\lambda'_{kl}$
present at the $\tilde{\phi}_{\overline{\mathcal{T}}}$ vertex will be
rotated by some $U^u$. The down quark field shown is a mass eigenstate
quark resulting from unitary the rotation, which we can interpret as a
linear combination of flavor eigenstates: $d_j = U^d_{jm}\, d'_{m}$,
with $j=1$; applying the same thinking to the up quark, we can also
write $u_k^T = u'^T_{p}\, U^{u\,T}_{pk}$, with $k=1$. To work out the
details of the rotations, we can start with the $d=5$ operator written
in terms of flavor states\footnote{Recall the scalars are both mass
and flavor eigenstates under the universal mass assumption. Also note
``$\lambda'$'' is again my name for the second generic coupling, and
the prime has nothing to do with basis; I will continue to use hats to
indicate rotated couplings.}, $\sum\nolimits_a x_a (\tilde{u}_i\,
\lambda^a_{im}\, d'_m) (u'_p \lambda'^a_{pl}\, \tilde{e}_l)$, where I
have expanded $C^L_{impl}$ in terms of its component couplings and
chosen the indices with the malice of forethought; now we can write
\begin{align}
  &\sum\limits_a x_a (\tilde{u}^T_i\, C^{-1} \lambda^a_{im}\, d'_{m})
  (u'^{\,T}_p\, \lambda'^a_{pl}\, C^{-1}\, \tilde{e}_l) \nonumber
\end{align}
\begin{align}
  = ~ &\sum\limits_a x_a (\tilde{u}^T_i\, C^{-1}
  \underbrace{\lambda^a_{im}\, U^{d\, \dagger}_{mj}}_{\displaystyle
  \equiv \hat{\lambda}^a_{ij}}\, \underbrace{U^d_{jn}\,
  d'_{n}}_{\displaystyle d_j}) (\underbrace{u'^{\,T}_p\,
  U^{u\,T}_{pk}}_{\displaystyle u_k^T}\, \underbrace{U^{u\,*}_{kq}\,
  \lambda'^a_{ql}}_{\displaystyle \equiv \hat{\lambda}'^a_{kl}}\,
  C^{-1}\, \tilde{e}_l). \nonumber
\end{align}
Using the new definitions for $\hat{\lambda}$, we can see that the
rotated coefficient $\widehat{C}^L$ corresponding to the expression in
eq.\,(\ref{eq:Cs}) has become
\begin{align}
  \widehat{C}^L_{ijkl} &= x_0 \hat{h}_{ij} \hat{h}_{kl} + x_1
  \hat{f}_{ij} \hat{f}_{kl} - x_3 \hat{h}_{ij} \hat{f}_{kl} + \dots
  \nonumber \\ 
  &= x_0 (h\, U_d^\dagger)_{ij} (U_u^* h)_{kl}\, +\, x_1
  (f\, U_d^\dagger)_{ij} (U_u^* f)_{kl}\, -\, x_3 (h\,
  U_d^\dagger)_{ij} (U_u^* f)_{kl}\, +\, \dots
\end{align}
Note that this version of $\widehat{C}^L$ is only valid for
$UDUE$-type operators with this orientation in the diagram, namely,
those with a scalar $\tilde{e}$; there is an analogous pair of
rotations for $UDUE$ with a scalar down and fermionic lepton, as well
as two each for $UDD\mathcal{N}$ and $U^{\mathcal{C}} D^{\mathcal{C}}
U^{\mathcal{C}} E^{\mathcal{C}}$, for a total of six possible schemes.

\paragraph*{From Quarks to Hadrons.} As mentioned above, the composite
hadrons $p$ and $K,\pi$ (in addition to the lepton) carry physical
momenta in the proton decay process, {\it not} the ``external'',
``physical'' quarks we see in the dressed operators above. Therefore
we are in need of calculating a factor like $\bra{\mathrm{M}}
(qq)q \ket{p}$, where M\:$= K, \pi$ is the final meson
state. More explicitly these objects will look like
\begin{align*}
  &\bra{K^+} \epsilon_{abc} (u^c s^b)_L\, d^a_L \ket{p} \\
  &\bra{K^0} \epsilon_{abc} (u^a s^c)_R\, u^b_L \ket{p} \\
  &\bra{\pi^0} \epsilon_{abc} (u^b d^c)_L\, u^a_R \ket{p} \\
  & \qquad \qquad \vdots
\end{align*}
Such matrix elements are calculated using either a three-point
function (for M, $p$, and the $(qq)q$ operator) on the lattice or
chiral Lagrangian methods; in either case, the result is determined in
part by a scaling parameter $\beta_H$ defined by $\bra{0} (qq)q
\ket{p(s)} = \beta_H P_L u_p (s)$, where $P_L$ is the left-chiral
projection matrix and $u_p (s)$ is the Dirac spinor for an incoming
proton of spin $s$. In principle $\beta_H$ is not necessarily the same
for cases where the quarks have different chiralities, but the values
usually differ only in sign, which is irrelevant when the entire
factor is squared in the decay width expression.

While lattice methods have advanced significantly since the early
years of SUSY GUT theory, there is still a substantial amount of
uncertainty present in the calculation of both $\beta_H$ and the
matrix element factors; some groups have even obtained contradictory
results when applying the two methods in the same work
\cite{gavela-king}. Some more recent works ({\it
e.g.}\,\cite{fukugita})~using more advanced statistics and larger
lattices seem to be converging on trustworthy answers, but it is still
normal to see results vary by factors of (1/2 - 5) for a single decay
mode from one method to the next, where the values for the matrix
elements themselves are $\mathcal{O}(10) \times \beta_H$. Thus I will
simply take the admittedly favorable approach of using
$\bra{\mathrm{M}} (qq)q \ket{p(s)} \sim \beta_H P u_p$ for all modes.

It is not uncommon to see values as low as $\beta_H = 0.003$ used in
other works calculating proton decay \cite{donoghue}, but while
calculated values have indeed varied as much as (0.003 - 0.65) over
the years \cite{fukugita}, the value is now most commonly found in the
range (0.006 - 0.03) \cite{claudson}, with a tendency to prefer
$\beta_H \sim 0.015$, as seen in \cite{fukugita}. Again, I will take a
slightly optimistic approach and use $\beta_H = 0.008$.

\begin{figure}[t]
\begin{center}
  \includegraphics{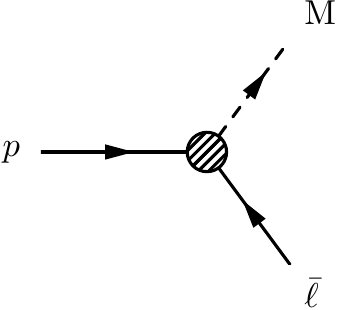} \caption[Proton decay to a meson
  and an anti-lepton.]{\footnotesize Proton decay to a meson and an
  anti-lepton for; the effective operator vertex contains hadronic and
  renormalization factors as well as the sum of all $d=6$ dressed
  operators contributing to the mode.} 
  \label{fig:pdecay}
\end{center}
\end{figure}

\paragraph*{The $p \rightarrow \mathrm{M} \bar\ell$ Effective Diagram
and the Decay Width of the Proton.} Ultimately it is a deceptively
simple two-body decay that I am calculating, as shown in Figure
\ref{fig:pdecay}. The corresponding decay width can be determined by
the usual phase-space integral expression:
\begin{equation}
  \Gamma = \frac{1}{2 M_p} \int \frac{\mathbf{d}^3
  \mathbf{p}}{(2\pi)^3\, 2 E_{\mathrm{M}}} \int \frac{\mathbf{d}^3
  \mathbf{p}}{(2\pi)^3\, 2 E_\ell}\, (2\pi)^4\; \delta^4(p_p -
  p_{\mathrm{M}} - p_\ell)\; \frac{1}{2}\, \sum\limits_s\; \lvert
  \mathscr{M} \rvert^{\,2}
\end{equation}
where in this case
\begin{equation}
  \frac{1}{2}\, \sum\limits_s\; \lvert \mathscr{M} \rvert^{\,2} =
  \frac{1}{2}\, \beta_H^2\, (A_L\,A_S)^2 \left( \lvert
  \mathscr{O}_{\widetilde{W}} \rvert^{\,2} + \lvert
  \mathscr{O}_{\tilde{h}} \rvert^{\,2} \right)\; \sum\limits_{s,s'}\;
  \lvert v^T_\ell (p_\ell,s)\, C^{-1}\, u_p (p_p,s') \rvert^{\,2}.
\end{equation}
\noindent The factors $A_L$ and $A_S$ arise to due the renormalization
of the $d=6$ dressed operators, from $M_p$ to $M_{SUSY}$ and
$M_{SUSY}$ to $M_U$, respectively; their values have been calculated
in the literature as $A_L = 0.4$ and $A_S = 0.9 \mbox{-} 1.0$
\cite{hisano}. The spinor factor can be evaluated with the usual trace
methods; in the rest frame of the proton where
$-\mathbf{p}_{\mathrm{M}} = \mathbf{p}_\ell \equiv \mathbf{p}$, and
utilizing $m_\ell^2 \ll \lvert \mathbf{p} \rvert^{\,2}$ (which is only
marginally valid for the muon but clearly so otherwise), the decay
width expression simplifies to
\begin{equation}
  \Gamma = \frac{1}{4\pi}\, \beta_H^2\, (A_L\,A_S)^2 \left( \lvert
  \mathscr{O}_{\widetilde{W}} \rvert^{\,2} + \lvert
  \mathscr{O}_{\tilde{h}} \rvert^{\,2} \right)\; \mathrm{p},
  \label{eq:gtotOs}
\end{equation}
where
\begin{equation}
  \mathrm{p} \equiv \lvert \mathbf{p} \rvert \simeq
  \frac{M_p}{2}\left(1 - \frac{m^2_{\mathrm{M}}}{M_p^2} \right).
\end{equation}
Note that p $\sim M_p/2$ for pion modes, but that value is
reduced by a factor of $\sim$\,25\% for kaon modes.

I now have all the pieces needed to write the working formulae for the
partial decay widths of the proton. Let me first define
$\mathrm{C}^{\mathcal{A}}$ as extended forms of the $C_{ijkl}$
by
\begin{align}
  \mathscr{C}_{\widetilde{W}}^{\mathcal{A}} =
  \mathrm{C}_{\widetilde{W}}^{\mathcal{A}} (qq)(q\ell) \nonumber \\
  \mathscr{C}_{\tilde{h}^\pm}^{\mathcal{A}} =
  \mathrm{C}_{\tilde{h}^\pm}^{\mathcal{A}} (qq)(q\ell) \nonumber \\
  \mathscr{C}_{\tilde{h}^0}^{\mathcal{A}} =
  \mathrm{C}_{\tilde{h}^0}^{\mathcal{A}} (qq)(q\ell),
\end{align}
so that these coefficients contain the $U^f/2$ or $y^f$ factors as well
as the $C_{ijkl}$ of the $\mathscr{C}^{\mathcal{A}}$ operators in
(\ref{eq:CWops})-(\ref{eq:Ch0ops}). Now I can easily translate an
operator expression like
\begin{equation}
  \mathscr{O}_{\widetilde{W}} (K^+ \bar{\nu}) \simeq 
  \left(\frac{i \alpha_2}{4\pi}\right) \frac{1}{M_\mathcal{T}}\,
  \left(\frac{M_{\widetilde{W}}}{m_{\tilde{q}}^2}\right)
  \{\mathscr{C}_{\widetilde{W}}^I + \mathscr{C}_{\widetilde{W}}^{I\!V}
  \}
  \label{eq:OKnuW}
\end{equation}
into a partial decay width statement,
\begin{equation}
  \Gamma_{\widetilde{W}} (p \rightarrow K^+ \bar{\nu}) \simeq
  \frac{1}{4\pi} \left(\frac{\alpha_2}{4\pi}\right)^2
  \frac{1}{M^2_{\mathcal{T}}} \left(\frac{M_{\widetilde{W}}}
  {m^2_{\tilde{q}}}\right)^2 \beta_H^2\,(A_L A_S)^2\, \mathrm{p}\:
  \lvert\, \mathrm{C}_{\widetilde{W}}^I +
  \mathrm{C}_{\widetilde{W}}^{I\!V}\, \rvert^{\,2}, 
  \label{eq:gKnuW}
\end{equation}
without losing either information or readability. Note though there is
still a ``black-box'' nature to the $\mathrm{C}^{\mathcal{A}}$ (it was
there in the $\mathscr{C}^{\mathcal{A}}$ operators as well), in that
without specifying the generation indices of the external $d_{j,i'}$
quarks, the sums in eqs.\,(\ref{eq:OKnuW}) and (\ref{eq:gKnuW}) could
just as easily apply to $\pi^+ \bar{\nu}$. Furthermore, there are at
least several channels present in each $\mathscr{C}^{\mathcal{A}}$
operator that contribute to any one mode, which are determined
uniquely by the generations of the internal sfermions in addition to
those of the external quarks.\footnote{Indeed I could have defined the
coefficients with six indices: $\mathrm{C}^{\mathcal{A}}_{ijklmn}$,
thereby creating a means of alleviating all degeneracy, but I don't
expect such information-dense objects to be so enlightening to
readers, especially since for most modes, at least the
Higgsino-dressed expression would devolve into an entire pageful of
terms corresponding to the individual channels.} If the reader wishes
to examine the decay widths at the full level of detail, he or she
should utilize these expressions along with the operators in
eqs.\,(\ref{eq:CWops})-(\ref{eq:Ch0ops}) and the diagrams in Appendix
\ref{fds}.

All remaining limitations aside, I can now present relatively compact
and intelligible expressions for the Wino- and Higgsino-dressed
partial decay widths of the proton for generic mode $p \rightarrow
\mathrm{M} \bar\ell$:
\begin{align}
  \label{eq:gammaW}
  \Gamma_{\widetilde{W}} (p \rightarrow \mathrm{M} \bar\ell) \simeq
  \frac{1}{4\pi} \left(\frac{\alpha_2}{4\pi}\right)^2
  \frac{1}{M^2_{\mathcal{T}}} \left(\frac{M_{\widetilde{W}}}
  {m^2_{\tilde{q}}}\right)^2 \beta_H^2\,(A_L A_S)^2\, \mathrm{p}\:
  \Big\lvert\! \sum\limits_{\mathcal{A} \in \mathrm{M} \bar\ell}\!
  \mathrm{C}_{\widetilde{W}}^{\mathcal{A}}\, \Big\rvert^{\,2} \\
  \Gamma_{\tilde{h}} (p \rightarrow \mathrm{M} \bar\ell) \simeq
  \frac{1}{4\pi} \left(\frac{1}{16\pi^2}\right)^2
  \frac{1}{M^2_{\mathcal{T}}} \left(\frac{\mu}
  {m^2_{\tilde{q}}}\right)^2 \beta_H^2\,(A_L A_S)^2\, \mathrm{p}\:
  \Big\lvert\! \sum\limits_{\mathcal{A} \in \mathrm{M} \bar\ell}\!
  \mathrm{C}_{\tilde{h}}^{\mathcal{A}} \Big\rvert^{\,2}. 
  \label{eq:gammah}
\end{align}
For the numerical analysis, I used the generic values $M_{\cal T} = 2
\!\times\! 10^{16}$\,GeV, $M_{\widetilde{W}} = \mu = 100$\,GeV, and
$m_{\tilde{q}} = 3$\,TeV. Also, let me repeat here that because of the
two SUSY Yukawa coupling factors in the
$\mathrm{C}_{\tilde{h}}^{\mathcal{A}}$, which always come in opposite
flavor,
\begin{equation}
  \Gamma_{\tilde{h}} \propto \left(\frac{1}{v_{\rm wk}^2
  \sin\beta\cos\beta} \right)^2 \sim
  \frac{\tan^2\beta}{v_{\rm wk}^4}. \nonumber
\end{equation}

Before moving on to the fermion sector fit results, let me remark that
because the Higgsinos vertices change the chiralities of the outgoing
fermions, there can be no interference between Wino- and
Higgsino-dressed diagrams, as suggested by eq.\,(\ref{eq:gtotOs});
however, since diagrams for the right-handed $C^R$ operators have
outgoing \emph{left-handed} fermions by the same Higgsino mechanism,
diagrams for $C^R$- and $C^L$-type operators with the same external
particles of matching chiralities \emph{do} interfere with each other,
and so all such contributions to a given mode do in fact go into the
same absolute-squared sum factor, as suggested by
eq.\,(\ref{eq:gammah}). 

\section{Fitting the Fermion Mass Matrices} \label{fit} Diagonalizing
the mass matrices given in eq.\,(\ref{eq:mass}), with the Yukawa
textures shown in (\ref{eq:y}), gives the GUT-scale fermion masses and
mixing angles for a given set of values for the mass matrix parameters
$h_{ij}$, $f_{ij}$, $r_i$, etc. In order to find the best fit to the
experimental data, I used the {\tt Minuit} tool library for Python
\cite{minuit,python} to minimize the sum of chi-squares for
the mass-squared differences $\Delta m_{21}^2$ (aka $\Delta
m_\odot^2$) and $\Delta m_{32}^2$ (aka $\Delta m_{\rm atm}^2$) and the
PMNS mixing angles in the neutrino sector as well as the mass
eigenvalues and CKM mixing angles in the charged-fermion sector.
Type-I and type-II seesaw neutrino masses were each fit independently,
so I report the results for each separately.

Note that throughout the analysis, I have taken $v_u = 117.8$\,GeV,
which is calculated with $\tan\beta = 55$ and for $v_{\rm wk}$ run to
the GUT scale \cite{das}. The corresponding value for the down-type
vev is $v_d = 2.26$\,GeV.

Threshold corrections at the SUSY scale are $\propto\tan\beta$, and so
should be large in this analysis \cite{poko}. The most substantial
correction is to the bottom quark mass, which is dominated by gluino
and chargino loop contributions; this correction also induces changes
to the CKM matrix elements involving the third generation. The
explicit forms of these corrections can be seen in a previous work on
a related model \cite{ddms}. Additionally, smaller off-diagonal
threshold corrections to the third generation parts of $\mathcal{M}_d$
result in small corrections to the down and strange masses as well as
further adjustments to the CKM elements. All such corrections can be
parametrized in the model by
\begin{align}
  \mathcal{M}'_d = \mathcal{M}_d + 
    \frac{r_1}{\tan\beta}\left(\begin{array}{ccc}
    0 & 0 & \delta V_{ub}\\ 0 & 0 & \delta V_{cb}\\
    \delta V_{ub} & \delta V_{cb} & \delta m_b \end{array}\right),
\end{align}
\smallskip
where $\mathcal{M}_d$ is given by eq.\,(\ref{eq:mass}). If I simply
take this augmented form for $\mathcal{M}_d$ as part of the model
input, the $\delta$ parameters are fixed by the mass matrix fitting,
which results in implied constraints on certain SUSY parameters and
the mass values that depend on them, namely, the Higgs and the light
stop and sbottom masses. This entire prescription and its implications
were considered in detail in \cite{ddms}, and in comparing to that
work, one can see that for large $\tan\beta$ and relatively small
threshold corrections, the resulting constraints on the Higgs and
squark masses are less interesting, so I will not consider them in
more detail for this analysis.

\subsection*{Fit Results for Type II Seesaw} If one takes the
$\overline{\mathbf{126}}$ SM-singlet vev $v_R \gsim 10^{17}$\,GeV
({\it i.e.},\;the GUT scale), and the triplet vev $v_L \sim 1$\,eV,
then the type-II contribution ($v_L$ term) in eq.\,(\ref{eq:neu})
dominates over the type-I contribution ($v_R$ term) by an average of
two orders of magnitude in the neutrino mass matrix; therefore
eq.\,(\ref{eq:neu}) reduces to
\begin{equation}
  {\cal M}_\nu \simeq v_L f
\end{equation}
Using this prescription, I find a fairly large parameter space for
which the sum of chi-squares is quite low, although some of the output
values, such as $\theta_{13}$ and the down and bottom masses, are
quite sensitive to the variation in the minima. This is problematic
for $\theta_{13}$ especially, since it is known to high experimental
precision \cite{dayabay}. Tables \ref{table:paramsII} and
\ref{table:fitII} display the properties of one of the more favorable
fits; Table \ref{table:paramsII} gives the values for the adjusted
model input parameters, and Table \ref{table:fitII} gives the
corresponding output values for the fermion parameters, with
experimentally measured values included for comparison. Note that the
down quark mass is seemingly a bit low, which seems to be a general
feature in this model, but I will discuss in the next section why this
is not a problem. The precise value of $v_L$ for this fit is
1.316\,eV, which I chose to fix the overall neutrino mass scale at
$m_3 \sim 0.05$\,eV.
\begin{table}[t]
\begin{center}
  \begin{tabular}{||c|c||c|c||}\hline\hline
    $M$ (GeV) &               106.6& $r_1/ \tan\beta$ &    0.014601\\ 
    $f_{11}$ (GeV) &      -0.045564& $r_2$ &              0.0090315\\
    $f_{12}$ (GeV) &       0.048871& $r_3$ &                 1.154 \\ 
    $f_{13}$ (GeV) &       -0.59148& $c_e$ &                -2.5342\\ 
    $f_{22}$ (GeV) &       -2.06035& $c_\nu$ &                  n/a\\     
    $f_{23}$ (GeV) &        -1.4013& $\delta m_b$ (GeV) &   -22.740\\
    $f_{33}$ (GeV) &       -1.40644& $\delta V_{cb}$ (GeV) & 1.2237\\
    $g_{12}$ (GeV) &       0.018797& $\delta V_{ub}$ (GeV) & 4.2783\\
    $g_{13}$ (GeV) &       -0.92510& & \\
    $g_{23}$ (GeV) &        -3.8353& & \\
    \hline\hline
  \end{tabular}
  \caption{\footnotesize Best fit values for the model parameters at the
  GUT scale with type-II seesaw. Note that $c_\nu$, which appears in
  the Dirac neutrino mass contribution to the type-I term, is not
  relevant for type-II.}
\label{table:paramsII}
\end{center}
\end{table}
\begin{table}[t]
\begin{center}
  \begin{tabular}{||c|c|c||c|c|c||}\hline\hline
    & best fit & exp value & & best fit & exp value\\ \hline
    $m_u$ (MeV) &       0.7172 &    $0.72^{+0.12}_{-0.15}$ &
    $V_{us}$ &          0.2245 &    $0.2243\pm 0.0016$\\
    $m_c$ (MeV) &       213.8 &    $210.5^{+15.1}_{-21.2}$ &
    $V_{ub}$ &          0.00326&    $0.0032\pm 0.0005$\\
    $m_t$ (GeV) &       106.8  &    $95^{+69}_{-21}$ &
    $V_{cb}$ &          0.0349 &    $0.0351\pm 0.0013$\\
    $m_d$ (MeV) &       0.8827 &    $1.5^{+0.4}_{-0.2}$ &
    $J \times 10^{-5}$ & 2.38 &    $2.2\pm 0.6$\\
    $m_s$ (MeV) &       34.04 &    $29.8^{+4.18}_{-4.5}$ &
    $\Delta m_{21}^2 / \Delta m_{32}^2$ & 0.03065& $0.0309\pm 0.0015$\\
    $m_b$ (GeV) &       1.209 &    $1.42^{+0.48}_{-0.19}$ &
    $\theta_{13}~(^\circ)$ & 9.057 &$8.88\pm 0.385$\\
    $m_e$ (MeV) &       0.3565 &    $0.3565^{+0.0002}_{-0.001}$ &
    $\theta_{12}~(^\circ)$ & 33.01 &$33.5\pm 0.8$\\
    $m_\mu$ (MeV) &     75.297 &    $75.29^{+0.05}_{-0.19}$ &
    $\theta_{23}~(^\circ)$ & 47.70 &$44.1\pm 3.06$\\
    $m_\tau$ (GeV) &    1.635 &    $1.63^{+0.04}_{-0.03}$ &
    $\delta_{\rm CP}~(^\circ)$ &   -7.506 & \\
    \hline
    & & & $\sum \chi^2$ & 6.0 & \\
    \hline\hline
  \end{tabular}
  \caption{\footnotesize Best fit values for the charged fermion
  masses, solar-to-atmospheric mass squared ratio, and CKM and PMNS
  mixing parameters for the fit with Type-II seesaw. The $1\sigma$
  experimental values are also shown for comparison \cite{das},
  \cite{pdg}, where masses and mixings are extrapolated to the GUT
  scale using the MSSM renormalization group equations (RGEs).  Note
  that the fit values for the bottom quark mass and the CKM mixing
  parameters involving the third generation shown here include the
  SUSY-threshold corrections} 
\label{table:fitII}
\end{center}
\end{table}

In order to calculate the $C_{ijkl}$ proton decay coefficients, as
well as for use in the neutrino mass matrix (\ref{eq:neu}), I needed
to determine the ``raw'' Yukawa couplings, $h,f,g$, from the
dimensionful couplings, $\tilde{h},\tilde{f},\tilde{g}$, of the mass
matrices given in eq.\,(\ref{eq:mass}), which are obtained directly
from the fit; to do so I need to extract the absorbed vev $v_u$ and
doublet mixing parameters $f(U^{\cal D}_{I\!J}, V^{\cal D}_{I\!J})$
mentioned in section \ref{model}. There is some freedom in the values
of those mixing elements from the viewpoint of this predominantly
phenomenological analysis, but they are constrained by both unitarity
and the ratios $r_i$ and $c_\ell$, which have been fixed by the
fermion fit. Again, see \cite{olddmm} for details, or see \cite{ddms}
for an example of such a calculation. The resulting dimensionless
couplings corresponding to this type-II fit are
\medskip
\begin{gather}
  h = \left( \begin{array}{ccc}
     0 & &\\ & 0 & \\ & & 1.207
  \end{array}\right) \qquad
  f = \left( \begin{array}{ccc}
    -0.00053748 &  0.00057649 & -0.0069772 \\
     0.00057649 & -0.024304 &  -0.016530 \\
    -0.0069772 &  -0.016530 & -0.0165906 \\
  \end{array} \right) \nonumber \\[4mm]
  g = i \left(\begin{array}{ccc}
     0          & 0.00033485 & -0.016480 \\
    -0.00033485 & 0          & -0.0683214 \\
     0.016480   & 0.0683214  &  0        \end{array} \right) 
\label{hfgII}
\end{gather}
\smallskip

\noindent Note that in addition to $f_{11} \sim f_{12} \sim 0$, this
fit satisfies $g_{12},f_{13} \ll 1$ as is desired for proton decay.

\subsection*{Fit Results for Type I Seesaw} If one instead takes $v_R
\lsim 10^{16}$\,GeV and $v_L \ll 1$\,eV, then the type-I
contribution is dominant over the type-II contribution, and
eq.\,(\ref{eq:neu}) becomes
\begin{equation}
  {\cal M}_\nu \simeq - {\cal M}_{\nu_D}\left(v_R f\right)^{-1}\left(
  {\cal M}_{\nu_D}\right)^T,
\end{equation}
In this case, initial searches again showed that certain output
parameters were quite sensitive to the input and were often in
contention with each other or with the {\it de-facto} upper bounds on
the $f_{ij}$ needed for proton decay. In the first cluster of minima
found by the fitting, the output values for one or more of charm mass,
bottom mass, or $\theta_{23}$ was much too small; furthermore, those
results came with odd, large tunings of certain input parameters, such
as $c_{e,\nu} \sim {\cal O}(100)$ or $\delta m_b > 40$\,GeV. The
addition of a small type-II correction to the neutrino matrix led me
to a new swath of parameter space, and ultimately I found a new
cluster of minima that did not require the correction. Table
\ref{table:paramsI} gives the values for the adjusted model input
parameters for one such pure type-I fit, and Table \ref{table:fitI}
gives the corresponding output values for the fermion parameters. Fits
in this swath of parameter space still have $c_\nu \sim 50$ and
$\delta m_b \sim 25$\,GeV, but this value for $c_\nu$, while slightly
strange, is quite readily accommodated by the doublet mixing
parameters, and such a value for the largest SUSY threshold correction
is actually quite moderate for large $\tan\beta$. The precise values
for the $\overline{\bf 126}$ vevs used in this fit are $v_L =
3.48$\,meV and $v_R = 1.21\! \times\! 10^{15}$\,GeV.

Note also that the top and strange masses are quite a bit lower than
in the type-II fit; however, note I have also quoted different
experimental values with which agreement is maintained. The
differences here come from an update to the work in \cite{das} in
determining two-loop MSSM RGEs for fermion masses. The update
\cite{bora} reports notably lower masses for all the quarks at
$\tan\beta = 55$ and $\mu = 2.0 \!\times\! 10^{16}$\,GeV, especially
for the up, down, strange, and top masses, due to updates in initial
values and methodology. Hence, one should not give the specific values
too much weight in such a fit, and I do not consider the reported
differences to be significant. This same thinking applies for the
type-II down mass value in Table \ref{table:fitII}.
\begin{table}[t]
\begin{center}
  \begin{tabular}{||c|c||c|c||}\hline\hline
    $M$ (GeV) &           76.10& $r_1/ \tan\beta$ &      0.024701\\
    $f_{11}$ (GeV) &   0.010130& $r_2$ &                  0.24414\\
    $f_{12}$ (GeV) &  -0.089576& $r_3$ &                  0.00600\\
    $f_{13}$ (GeV) &    0.93973& $c_e$ &                  -3.3279\\
    $f_{22}$ (GeV) &     0.8659& $c_\nu$ &                 45.218\\
    $f_{23}$ (GeV) &     1.4884& $\delta m_b$ (GeV) &     -28.000\\
    $f_{33}$ (GeV) &     3.5495& $\delta V_{cb}$ (GeV) & -0.84394\\
    $g_{12}$ (GeV) &    0.20048& $\delta V_{ub}$ (GeV) &  0.51486\\
    $g_{13}$ (GeV) &    0.05352& & \\
    $g_{23}$ (GeV) &    0.35153& & \\
    \hline\hline
  \end{tabular}
  \caption{\footnotesize Best fit values for the model parameters at the
  GUT scale with type-I seesaw.}
\label{table:paramsI}
\end{center}
\end{table}
\begin{table}[t]
\begin{center}
  \begin{tabular}{||c|c|c||c|c|c||}\hline\hline
    & best fit & exp value & & best fit & exp value\\ \hline
    $m_u$ (MeV) &       0.72155&    $0.72^{+0.12}_{-0.15}$ &
    $V_{us}$ &          0.2240 &    $0.2243\pm 0.0016$\\
    $m_c$ (MeV) &        212.2 &    $210.5^{+15.1}_{-21.2}$ &
    $V_{ub}$ &          0.00310&    $0.0032\pm 0.0005$\\
    $m_t$ (GeV) &        76.97 &    $80.45^{+2.9\,*}_{-2.6}$ &
    $V_{cb}$ &          0.0352 &    $0.0351\pm 0.0013$\\
    $m_d$ (MeV) &        1.189 &    $0.930\pm 0.38^*$ &
    $J \times 10^{-5}$ & 2.230 &    $2.2\pm 0.6$\\
    $m_s$ (MeV) &        20.81 &    $17.6^{+4.9\,*}_{-4.7}$ &
    $\Delta m_{21}^2 / \Delta m_{32}^2$ & 0.0309 & $0.0309\pm 0.0015$\\
    $m_b$ (GeV) &        1.278 &    $1.24\pm 0.06^*$ &
    $\theta_{13}~(^\circ)$ & 8.828 &$8.88\pm 0.385$\\
    $m_e$ (MeV) &       0.3565 &    $0.3565^{+0.0002}_{-0.001}$ &
    $\theta_{12}~(^\circ)$ & 33.58 &$33.5\pm 0.8$\\
    $m_\mu$ (MeV) &      75.29 &    $75.29^{+0.05}_{-0.19}$ &
    $\theta_{23}~(^\circ)$ & 41.76 &$44.1\pm 3.06$\\
    $m_\tau$ (GeV) &     1.627 &    $1.63^{+0.04}_{-0.03}$ &
    $\delta_{\rm CP}~(^\circ)$ &   -46.3 & \\
    \hline
    & & & $\sum \chi^2$ & 1.75& \\
    \hline\hline
  \end{tabular}
  \caption{\footnotesize Best fit values for the charged fermion
  masses, solar-to-atmospheric mass squared ratio, and CKM and PMNS
  mixing parameters for the fit with Type-I seesaw. The $1\sigma$
  experimental values are shown \cite{das} ($^*$\,-\,\cite{bora}),
  \cite{pdg}; masses and mixings are extrapolated to the GUT scale
  using the MSSM RGEs. Note that again that pertinent fit values
  include threshold corrections.} 
  \vspace{-4mm}
\label{table:fitI}
\end{center}
\end{table}

Again I need to determine the raw Yukawa couplings for proton decay
analysis. The resulting couplings corresponding to this type-I fit
are \smallskip
\begin{gather}
  h = \left( \begin{array}{ccc}
     0 & &\\ & 0 & \\ & & 1.6152
  \end{array}\right) \qquad
  f = \left( \begin{array}{ccc}
     0.0001623 &  -0.00143525 &  0.01505699 \\
    -0.00143525 &  0.01387415 &  0.02384774 \\
     0.01505699 &  0.02384774 &  0.05687217
  \end{array} \right) \nonumber \\[4mm]
  g = i \left(\begin{array}{ccc}
     0          & 0.0068081  & 0.0018175 \\
    -0.0068081  & 0          & 0.0119376 \\
    -0.0018175  &-0.0119376  & 0        \end{array} \right) 
\label{eq:hfgI}
\end{gather} 
\smallskip

\noindent Here, we still see $f_{11} \sim 0$, but each of $f_{12}$,
$f_{13}$, and $g_{12}$ is larger by an order of magnitude than in the
type-II case, which is thought to be unfavorable for proton decay. At
the same time, $g_{13}$ and $g_{23}$ are smaller by an order of
magnitude, so it is not clear that the net benefit lost is
substantial. In the end, a different distinction will give way to
success for this type-I fit; I will discuss those details in the next
section.

\section{Results of Calculating Proton Partial Lifetimes} \label{pfit}
In order to give an actual number for any decay width, in addition to
choosing representative values for the triplet, sfermion, and Wino or
Higgsino masses, I also need values for the $x_i$ and $y_i$ triplet
mixing parameters in order to calculate the $C_{ijkl}$ values. Recall
that the {\bf 10} mass parameter $x_0$ must be ${\cal O}(1)$ to allow
the SUSY Higgs fields to be light; the remaining mixing parameters are
functions of many undetermined GUT-scale masses and couplings found in
the full superpotential for the heavy Higgs fields, the details of
which can be seen in \cite{aulgarg}. There are nearly as many of those
GUT parameters as there are independent $x$s and $y$s, so it is not
unreasonable to simply treat the latter as free parameters.

Ideally, one would find that the width for any particular mode would
be essentially independent of those parameter values, {\it i.e.}, that
for arbitrary choices $0 < |x_i|,|y_i| < 1$, devoid of unlucky
relationships leading to severe enhancements, all mode lifetimes would
be comfortably clear of the experimentally determined lower limits,
given in Table \ref{table:explims}. The reality is quite bleak in
comparison. For a typical GUT model, if the proton decay lifetimes can
be satisfied at all, one is required to choose $x$ and $y$ values very
carefully such that either individual $C$s or $\Big\lvert\! \sum
\mathrm{C}^{\mathcal{A}} \Big\rvert$ are small through cancellations
among terms. These tunings may need to be several orders of magnitude
in size ({\it e.g.}, $\mathrm{C}^{\cal A} = -\mathrm{C}^{\cal B} +
{\cal O}(10^{-3})$), and many such relationships may be needed. 
\begin{table}[t]
\begin{center}
  \begin{tabular}{||c|c||}\hline\hline
    decay mode & $\tau$ exp lower limit (yrs) \\ \hline 
    $p \rightarrow K^+ \bar\nu$   &  $6.0 \!\times\! 10^{33}$ \\
    $p \rightarrow K^0 e^+$       &  $1.0 \!\times\! 10^{33}$ \\
    $p \rightarrow K^0 \mu^+$     &  $1.3 \!\times\! 10^{33}$ \\
    $p \rightarrow \pi^+ \bar\nu$ &  $2.7 \!\times\! 10^{32}$ \\
    $p \rightarrow \pi^0 e^+$     &  $1.3 \!\times\! 10^{34}$ \\
    $p \rightarrow \pi^0 \mu^+$   &  $1.0 \!\times\! 10^{34}$ \\
    \hline\hline
  \end{tabular}
  \caption{\footnotesize Experimentally determined lower limits
  \cite{babuexp} on the partial lifetimes of dominant proton decay
  modes considered in this work.}
\label{table:explims}
\end{center}
\end{table}

The Yukawa textures shown in eq.\,(\ref{eq:y}) are intended to
naturally suppress the values of some crucial $C$ values so that the
need for such extreme tuning is alleviated. In order to test the
ansatz, I ``simply'' needed to find a set of values for the mixing
parameters yielding partial decay widths that satisfy the experimental
constraints; the difficulty in determining those values inversely
corresponds to success of the ansatz. If the ansatz does indeed work
optimally, I should be able to choose arbitrary $x_i$ and $y_i$ values
as suggested above.  Realistically though, the authors of
\cite{dmm1302} and I expected some searching for a valid region of
parameter space to be required.

To perform that search, I designed a second Python program to find
maximum partial lifetimes based on user-defined mixing values as well
as the raw Yukawa couplings fixed by the fermion sector fitting.
Parameter values are defined on a per-trial basis for any number of
trials. I started with the most optimistic case by generating random
initial values for $x_i$ and $y_i$ (but $x_0 \sim 1$ fixed), with the
decay width for $K^+ \bar\nu$ minimized by adjusting those values in
each trial. The minimization was again performed using the {\tt
Minuit} tool library.

The search based on fully random initial values was unsuccessful, in
that the $K^+ \bar\nu$ mode lifetime consistently fell in the $10^{31
\mbox{-} 32}$\,year-range for the type-II solution and was 
\noindent typically $\sim\!1 \!\times\! 10^{33}$\,years for the type-I
case;\footnote{The {\tt Minuit} tool used, Migrad, works using a local
gradient-based algorithm, so that in large parameter spaces, initial
values are crucial in locating global minima.} at the same time
however all five other modes in question were usually near or above
their respective limits for those same arbitrary mixing values.  Hence
it was clear that, even with the $K^+ \bar\nu$ mode failure, that the
ansatz was having the desired effect to some extent. Also, note that
this type-I solution for $K^+ \bar\nu$ was short of the limit by only
about a factor of five.  This is surprising since the type-I-based
Yukawas reported in eq.\,(\ref{eq:hfgI}) fell short of meeting the
ansatz criteria. Given the differing behaviors of the two solutions, I
will report the remaining details in separate subsections once again.

\subsection*{Proton Partial Lifetimes for Type II Seesaw } To further
explore the properties of the ``default behavior'' of the lifetime
values in the model, I considered the case in which $x_0 \sim 1$ and
all other $x_i$ and $y_i$ are set to zero; one can see this case as
defining a baseline for the partial lifetimes, in that any $x_0$
terms in the $C$s not suppressed by the Yukawa textures are
necessarily large, and whereas problematic contributions from some
other $x_k$ with $k\neq 0$ may be suppressed simply by setting $x_k
\ll 1$, the $x_0$ contributions can be mitigated {\it only} through
cancellation. 

\begin{table}[t]
\begin{center}
  \begin{tabular}{||c|c|c||}\hline\hline
    decay mode & baseline for $\tau$ (yrs) & baseline in ref.
    \cite{altarelli} (yrs) \\ \hline 
    $p \rightarrow K^+ \bar\nu$   &  $8.29 \!\times\! 10^{31}$ &  
        $6.38 \!\times\! 10^{28}$ \\
    $p \rightarrow K^0 e^+$       &  $9.73 \!\times\! 10^{34}$ &
        $2.52 \!\times\! 10^{30}$ \\
    $p \rightarrow K^0 \mu^+$     &  $5.68 \!\times\! 10^{33}$ &
        $6.15 \!\times\! 10^{29}$ \\
    $p \rightarrow \pi^+ \bar\nu$ &  $4.25 \!\times\! 10^{33}$ &
        $4.45 \!\times\! 10^{29}$ \\
    $p \rightarrow \pi^0 e^+$     &  $1.08 \!\times\! 10^{36}$ &
        $3.90 \!\times\! 10^{30}$ \\
    $p \rightarrow \pi^0 \mu^+$   &  $6.45 \!\times\! 10^{34}$ &
        $6.00 \!\times\! 10^{29}$ \\
    \hline\hline
  \end{tabular}
  \caption{\footnotesize Hypothetical baseline partial lifetimes
  determined using type-II solution Yukawas and $x_0 = 0.95$ with all
  other $x_i,y_i = 0$. For comparison, I give the analogous results
  for calculation using type-II Yukawas from the 2010 paper by
  Alterelli and Blankenburg \cite{altarelli}, which use general Yukawa
  texture.  Note in comparing with Table \ref{table:explims} that for
  our model, only the $K^+ \bar\nu$ mode fails to satisfy the lower
  limit, while all modes are well below the limits for the model in
  \cite{altarelli}.}
\label{table:baselineII}
\end{center}
\end{table}

The corresponding baseline lifetimes for the dominant modes in the
type-II case are given in Table \ref{table:baselineII}. One can see
that the $K^+ \bar\nu$ mode decay width must be lowered by two orders
of magnitude through cancellation of $x_0$ terms by the others.  Since
it is $\lvert \mathrm{C} \rvert^{\,2}$ that appears in the decay width
expressions, the needed cancellation amounts to an ${\cal O}(10^{-1})$
tuning among the ${\rm C}^{\cal A}$ factors.  Furthermore, as it would
be equally unnatural to see $x_k \ll 1$ for all $k\neq 0$, one should
expect ${\cal O}(1)$ cancellations to be present anyway; therefore,
the needed ``tuning'' is little more than a very ordinary restriction
of parameter space.

In order to elucidate the significance of the improvement created by
the Yukawa ansatz, let us consider the outcome of this baseline
calculation for a case with more general Yukawa texture. The model
from a 2010 paper by G. Altarelli and G. Blankenburg \cite{altarelli}
has the same {\bf 10}-{\bf 126}-{\bf 120} Yukawa structure but with
general $h$ and $g$ as in eq.\,(\ref{eq:y0}) and a tri-bimaximal $f$
having no hierarchical texture.\footnote{This model has already been
ruled out due to $\theta_{13} \sim 6\mbox{-}7^\circ$ typical of
tri-bimaximal models.} Using the parameters reported to give a
successful fermion fit in the work (\emph{see footnote}), I obtain the
baseline results shown in the final column of Table
\ref{table:baselineII}. One can see here that lifetimes for all modes
are far below the experimental limits, by factors of ${\cal
O}(10^{3\mbox{-}5})$; hence cancellation among the ${\rm C}^{\cal A}$
factors must be ${\cal O}(10^{-2\mbox{-}4})$. Such sensitive
relationships among these factors are considerably more restrictive
than the result from our model, and, in the absence of some new
symmetry, there is no good explanation for those restrictions.

\begin{figure}[t]
        \includegraphics[width=7.8cm]{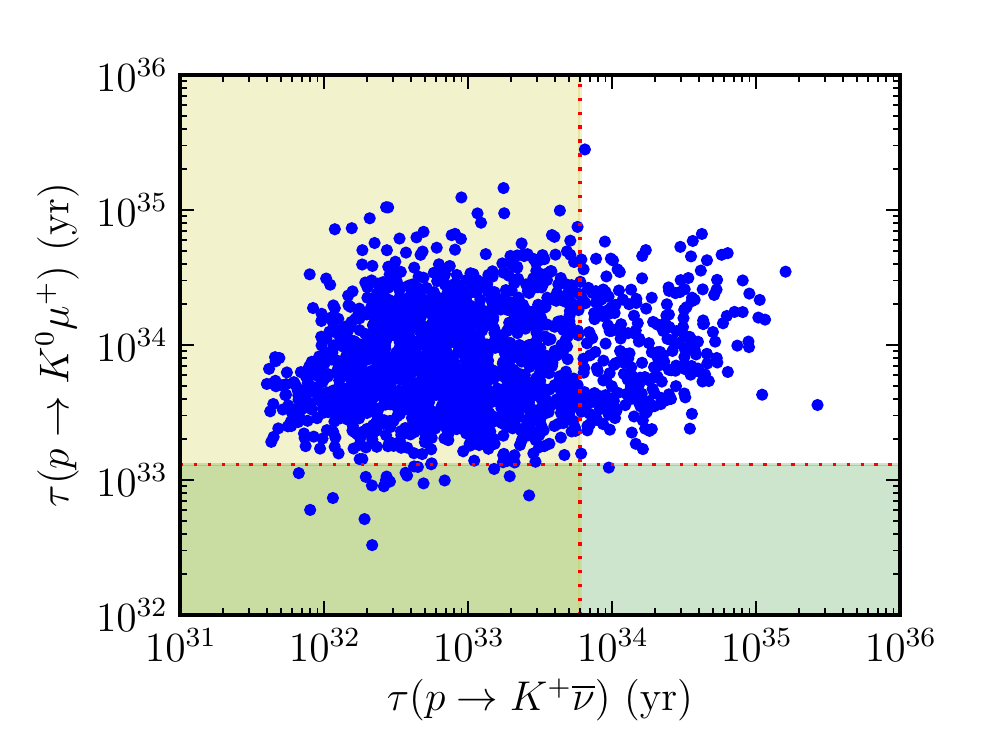}
        \includegraphics[width=7.8cm]{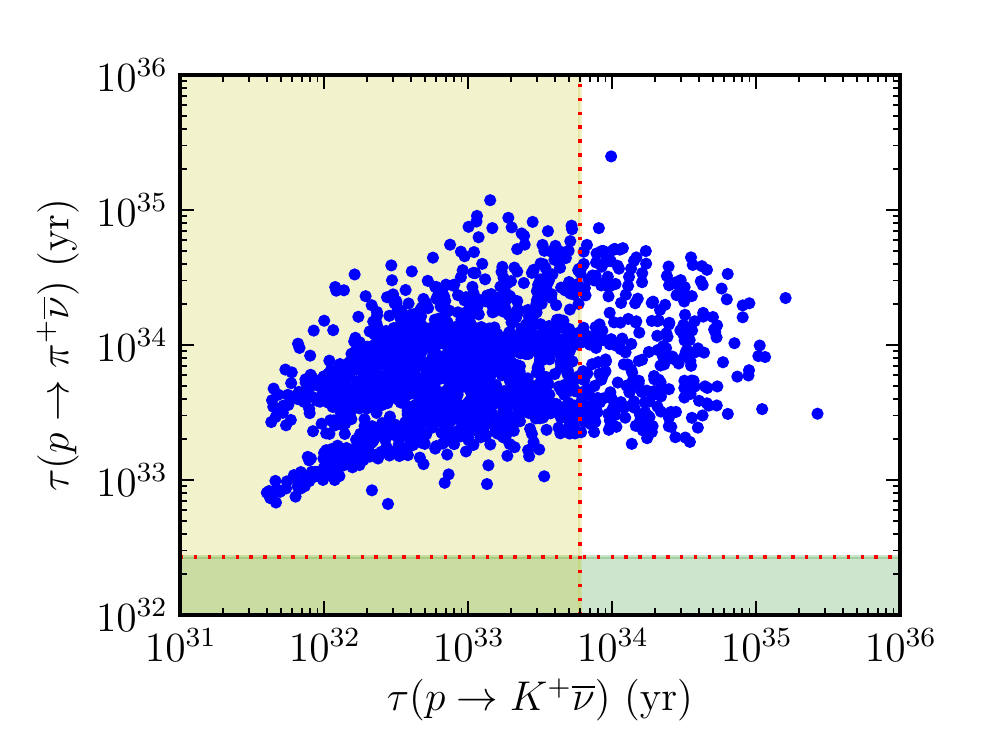}\\
        \includegraphics[width=7.8cm]{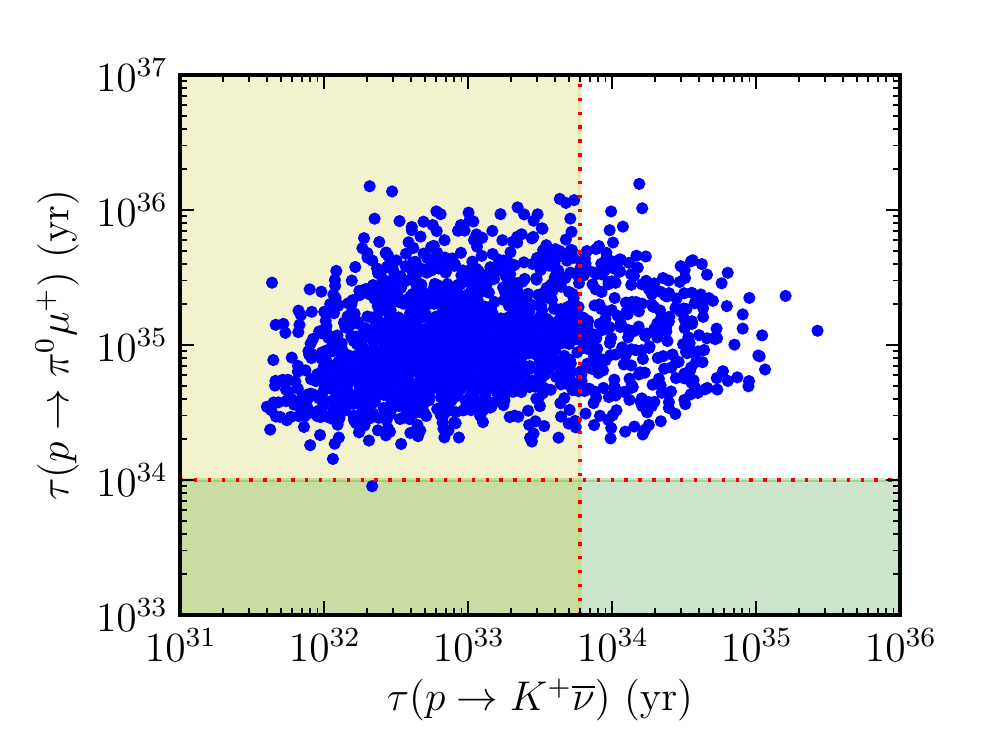}
        \includegraphics[width=7.8cm]{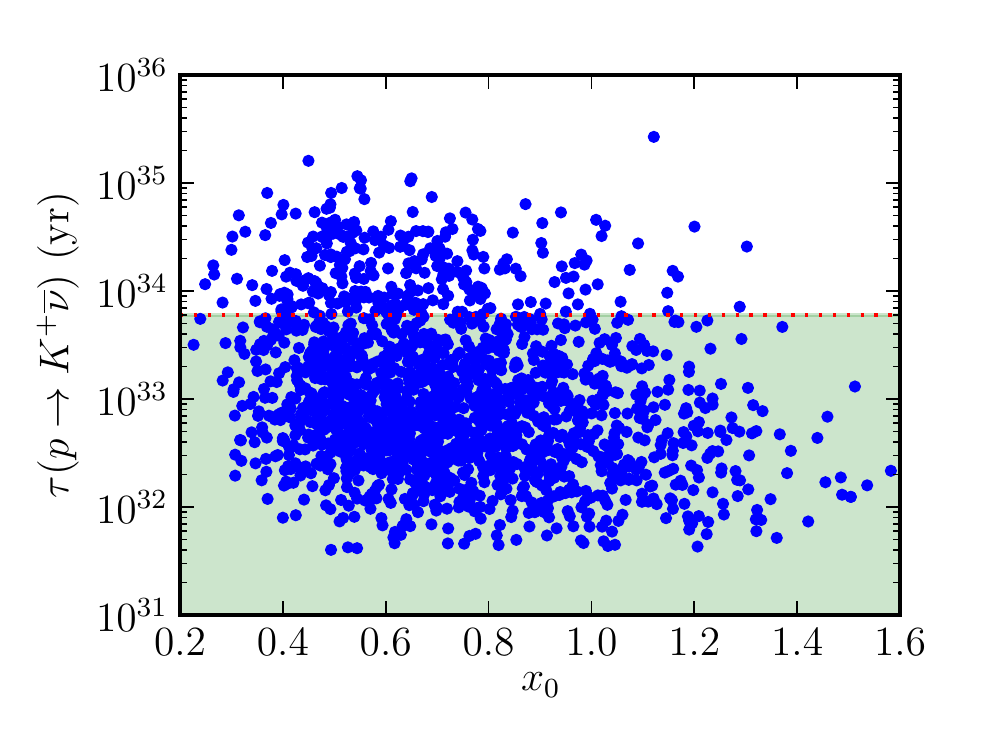}
        \caption{\footnotesize Comparisons of $K^+ \bar\nu$ partial
        lifetime to those of other dominant modes in the model, and
        that lifetime as a function of the {\bf 10} mass parameter
        $x_0$, for the type-II case. Note the unsurprising preference
        for smaller $x_0$.}
	\label{fig:KnuplotsII}
\end{figure}

In order to locate an area of mixing parameter space which yields a
sufficient $K^+ \bar\nu$ lifetime, I wrote a supplementary Mathematica
code to search for minima among strongly abridged versions of
$\lvert\, \mathrm{C}_{\widetilde{W}}^I +
\mathrm{C}_{\widetilde{W}}^{I\!V}\, \rvert$ and $\lvert\,
\mathrm{C}_{\tilde{h}^\pm}^{I\!V}\, \rvert$ that contribute to the
decay width.\footnote{$\mathrm{C}_{\tilde{h}^\pm}^{I\!I\!I}$ and
$\mathrm{C}_{\tilde{h}^0}^{I\!I\!I}$ cancel identically for all
contributing channels of both the $K^+ \bar\nu$ and $\pi^+ \bar\nu$
modes.} Specifically I started with $x_0$ terms only, corresponding to
the baseline case, and then iteratively added back the largest
contributions one by one while readjusting the initial values each
time. Once all of the most important terms were present, I took the
resulting mixing parameters as my initial values in the Python code.
The resulting minimization gave a large percentage of trials with all
six modes exceeding the lifetime bounds.

\begin{figure}[t]
        \includegraphics[width=7.8cm]{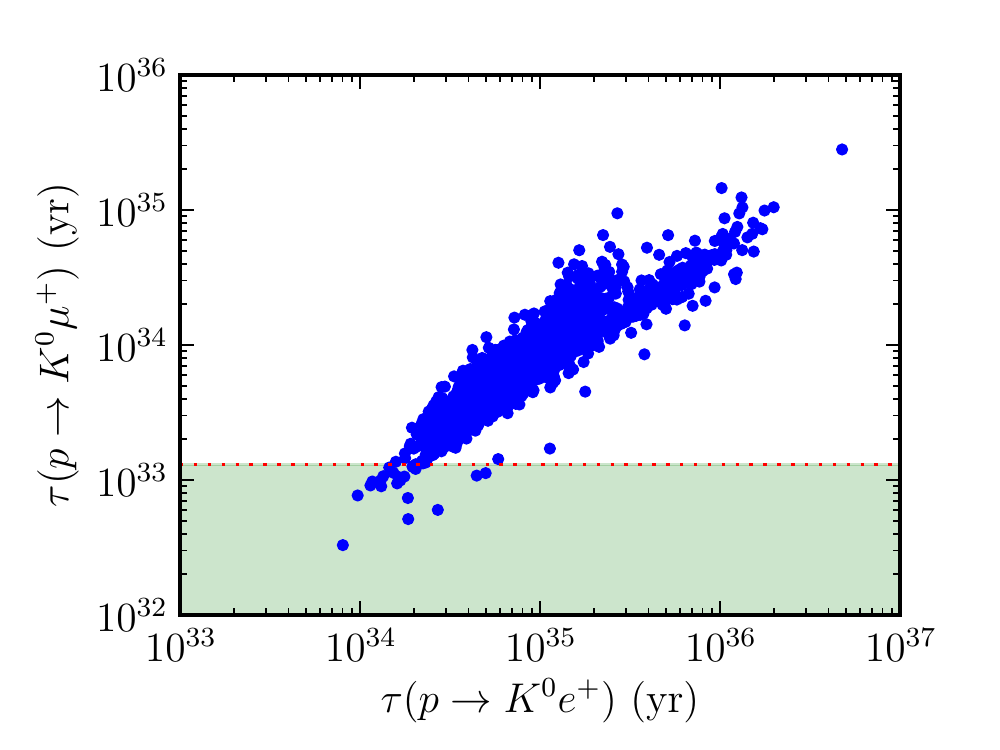}
        \includegraphics[width=7.8cm]{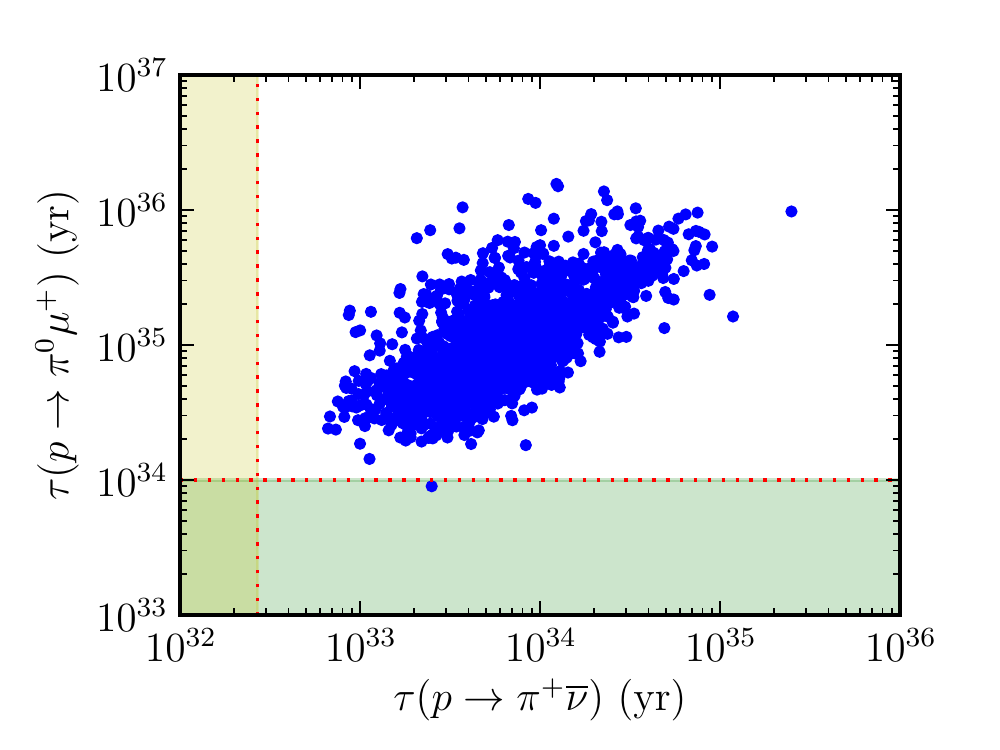}
        \caption{\footnotesize Comparisons of partial lifetimes among
        highly-correlated sub-dominant modes in the model for the
        type-II case.}
	\label{fig:otherplotsII}
\end{figure}

With an allowed region of parameter space found, I expanded my
searches to include a wider range of values for $x_0$. Using six
different ``seeds'' for parameter values, all of which give every mode
sufficient with $\tau(K^+ \bar\nu)$ roughly twice the experimental
bound, I created a large number of trials for which the initial values
were distributed normally around the seed values and with large
standard deviations. The resulting data for such a search is shown in
scatter plots below. Figure \ref{fig:KnuplotsII} gives the
relationships between the $K^+ \bar\nu$ mode and other representative
modes and also the distribution of $K^+ \bar\nu$ lifetime for varying
$x_0$. Figure \ref{fig:otherplotsII} shows the relationships between
other more closely correlated modes for completeness.

Note the strong correlation between $\pi^+ \bar\nu$ and $\pi^0 \mu^+$,
which are related by isospin, and the extreme correlation between $K^0
e^+$ and $K^0 \mu^+$. The latter is due to a manifestation of the
hierarchical nature of the Yukawas in the $C_{ijkl}$, as well as minor
features such $f_{11} \sim f_{12}$; similar structure is present in
the $y^f$ and $U^f$, which tend to also have $11 \sim 12$ or $11 \ll
12$; these properties result in a straightforward scaling under the
replacement $l: 1 \rightarrow 2$. Furthermore, the same relationship
is present between $\pi^0 e^+$ and $\pi^0 \mu^+$. These relationships
imply that the remaining plots I omitted differ only trivially from
the representatives present.

I also performed simple scans in search of a maximum value for
$\tau(K^+ \bar\nu)$, as well as taking note of any especially large
values in the previous searches. While there does not seem to be any
analytically-enforced maximum present in the model, I did consistently
find that $\tau > 10^{35}$\,years was extremely rare, and I never saw
a value higher than $\sim 6\!\times\! 10^{35}$\,yr. Given those
findings, combined with the apparent smallness of the swath of
parameter space yielding the above results and the low likelihood of a
more global minimum based on my search methods, I believe that
$\tau(K^+ \bar\nu) \gsim 10^{36}$\,yr is statistically infeasible in
this model for type-II seesaw. If such a value does exist, it is
likely contained in a vanishingly small area of allowed parameter
space and accomplished through truly extreme tuning. Therefore I will
take $10^{36}$\,years as a {\it de facto} upper limit on $\tau(K^+
\bar\nu)$ for the type-II case, which will not be accessible by
Hyper-K and similar experiments \cite{hyperk,dune} in the near future,
but should nonetheless allow the model to be tested eventually.

The other modes of course have similar limits, but it would seem that
all the others are substantially higher and thus either far beyond the
reach of the forthcoming experiments or beyond the contributions from
gauge boson exchange, if not both, with the possible exception of
$\tau(\pi^+ \bar\nu)$, which is rather highly correlated with $K^+
\bar\nu$ in this model. Determining that value is tricky though
because if I simply maximize the $\pi^+ \bar\nu$ mode, then the $K^+
\bar\nu$ mode will be below its bound; thus, there is some question as
to how one defines the maximization.

\subsection*{Proton Partial Lifetimes for Type I Seesaw } I begin
again by examining the same baseline case for the partial lifetimes,
with $x_0 \sim 1$ and all other $x_i,y_i = 0$. The resulting values
for the dominant modes in the type-I case are given in Table
\ref{table:baselineI}. Here we see a much more favorable situation, in
that even the $K^+ \bar\nu$ mode decay width is sufficient, and in
fact the other modes exceed the bounds by 2-4 orders of magnitude.
Hence we expect that virtually all solutions will be adequate for
modes other than $K^+ \bar\nu$, and as long as there is no {\it
enhancement} due to (de)tuning among the ${\rm C}^{\cal A}$ factors,
that mode will be adequate as well.

This is of course a remarkable improvement over traditional models,
yet it seems to contradict our expectations given then properties of
the fermion fit. Why then is the model successful? There are two
primary reasons, both of which are quite subtle. The first reason is
that the smaller values for $g_{13}$ and $g_{23}$ seen in
eq.\,(\ref{eq:hfgI}) do in fact improve the situation, as I
suggested, while the larger $f_{12}$ and $g_{12}$ seem to have less
impact. Since $M\;(h_{33})$ is such an extremely dominant factor in
the Yukawas, it is generally the case that contributions involving
third generation are larger and more important than the others.

\begin{table}[t]
\begin{center}
  \begin{tabular}{||c|c||}\hline\hline
    decay mode & baseline for $\tau$ (yrs) \\ \hline 
    $p \rightarrow K^+ \bar\nu$   &  $7.87 \!\times\! 10^{33}$ \\
    $p \rightarrow K^0 e^+$       &  $5.93 \!\times\! 10^{35}$ \\
    $p \rightarrow K^0 \mu^+$     &  $2.45 \!\times\! 10^{35}$ \\
    $p \rightarrow \pi^+ \bar\nu$ &  $2.37 \!\times\! 10^{36}$ \\
    $p \rightarrow \pi^0 e^+$     &  $6.11 \!\times\! 10^{38}$ \\
    $p \rightarrow \pi^0 \mu^+$   &  $2.27 \!\times\! 10^{38}$ \\
    \hline\hline
  \end{tabular}
  \caption{\footnotesize Hypothetical baseline partial lifetimes
  determined using type-I solution Yukawas and $x_0 = 0.95$ with all
  other $x_i,y_i = 0$. Note in comparing with Table \ref{table:explims} 
  that all modes satisfy the lower limits, and most do so by several
  orders of magnitude.}
  \vspace{-2mm}
\label{table:baselineI}
\end{center}
\end{table}

The second reason is even more unexpected, to the point that it was
not even examined in the preceding works on this ansatz. The unitary
matrices $U^f$ for the charged fermions are generally $\sim 1$, just
as one would expect, given the texture of CKM. This model is no
exception, with off-diagonal terms generally ${\cal O}(10^{-1 \mbox{-}
3})$; however, with such sparse or hierarchical (flavor basis) Yukawas
due to the ansatz, these ``small'' off-diagonal elements lead to
``small'' rotations of $h,f,g$ resulting in relatively substantial
chanes to the textures of $\hat{h},\hat{f},\hat{g}$. Especially
noteworthy are the changes in $h \rightarrow \hat{h}$, where some
previously-zero off-diagonal elements are replaced by the same ${\cal
O}(10^{-1 \mbox{-} 3})$ values seen in the $U^f$.

In light of the surprising non-triviality of the basis rotations, if
we compare $U^{u,d}$ for the type-I case: 
\begin{align}
  U^u = &\left( \begin{array}{ccc}
    0.994 & -0.1085 + 0.0057i &  0.00298 + 10^{-5}i \\
    0.1084 + 0.0057i &  0.994 &  0.0047 + 10^{-5}i  \\
   -0.0035 - 10^{-5}i & -0.0044 + 10^{-5}i &  0.99998
  \end{array} \right) \nonumber
\end{align}
\begin{align}
  U^d = &\left(\begin{array}{ccc}
    0.967 & -0.1087 + 0.2309i & 0.00175 + 0.001175i \\
    0.1086 + 0.2308i & 0.966  & 0.03935 + 0.00690i \\
   -0.0076 - 0.0072i & -0.0381 + 0.00613i & 0.9992
  \end{array} \right), 
\label{eq:UudI}
\end{align}
\smallskip
to those for the type-II case:
\begin{align}
  U^u = &\left( \begin{array}{ccc}
    0.972 &  0.2098 - 0.1044i & -10^{-5} - 0.010i \\
   -0.210 - 0.1043i &  0.971 & -0.00012 - 0.0414i  \\
   -0.0043 - 0.001i & -0.001 - 0.0423i &  0.999
  \end{array} \right) \nonumber
\end{align}
\begin{align}
  U^d = &\left(\begin{array}{ccc}
    0.9998 & 0.00633 - 0.0095i &  0.00765 - 0.01117i \\
   -0.00708 - 0.0095i &  0.9983 & 0.03386 - 0.04514i\\
   -0.00785 - 0.01054i & -0.03401 - 0.04514i &  0.9983        
  \end{array} \right), 
\label{eq:UudII}
\end{align}
we see that the off-diagonal entries are the same size or smaller for
the type-I case in every entry except $U^d_{12}$; furthermore,
several of the elements involving the third generation are smaller by
an order of magnitude. These differences may seem rather benign, but
in fact each of these slightly suppressed values individually
translates into a factor of 10 suppression in most of the dominant
$C$s, which all tend to involve third generation elements. In some
cases two or even three such suppressions may affect a single ${\rm
C}^{\cal A}$ factor. The squaring of factors in the decay width then
gives suppressions of generally 2-4 orders of magnitude in the
lifetimes, which is precisely what one can see when comparing Tables
\ref{table:baselineII} and \ref{table:baselineI}.
\begin{figure}[t]
        \includegraphics[width=7.8cm]{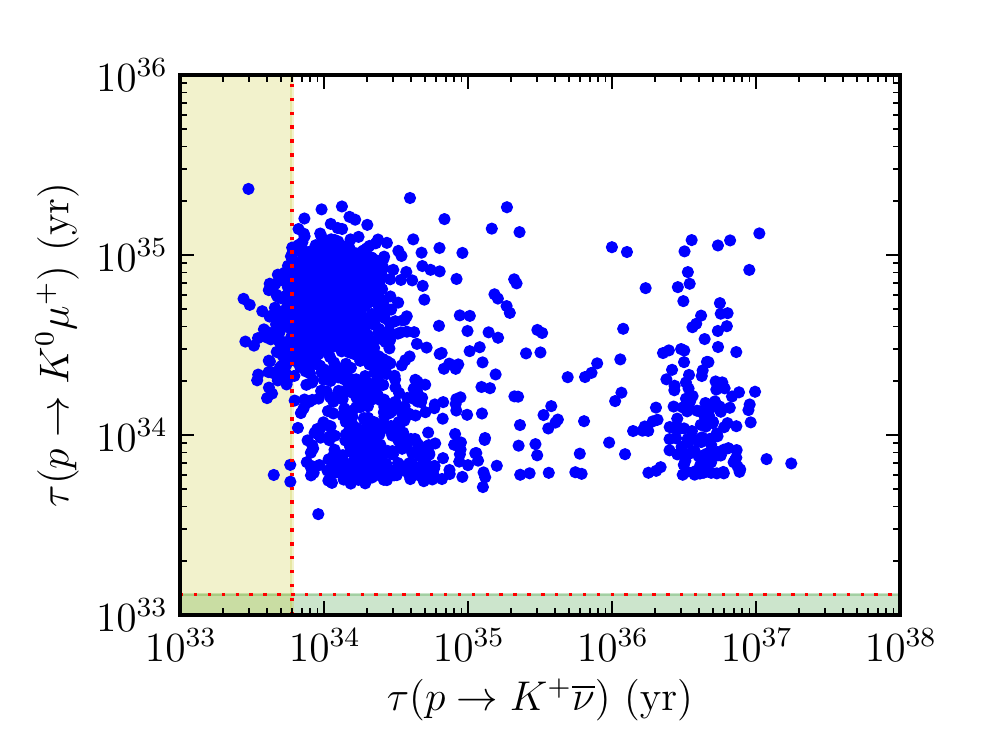}
        \includegraphics[width=7.8cm]{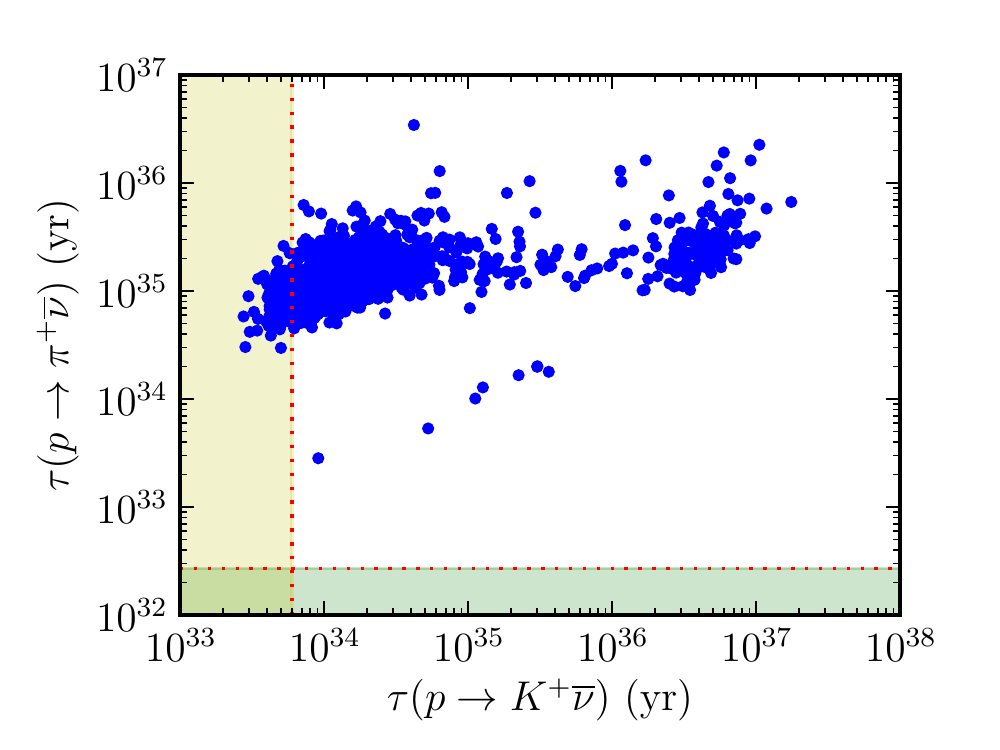}\\
        \includegraphics[width=7.8cm]{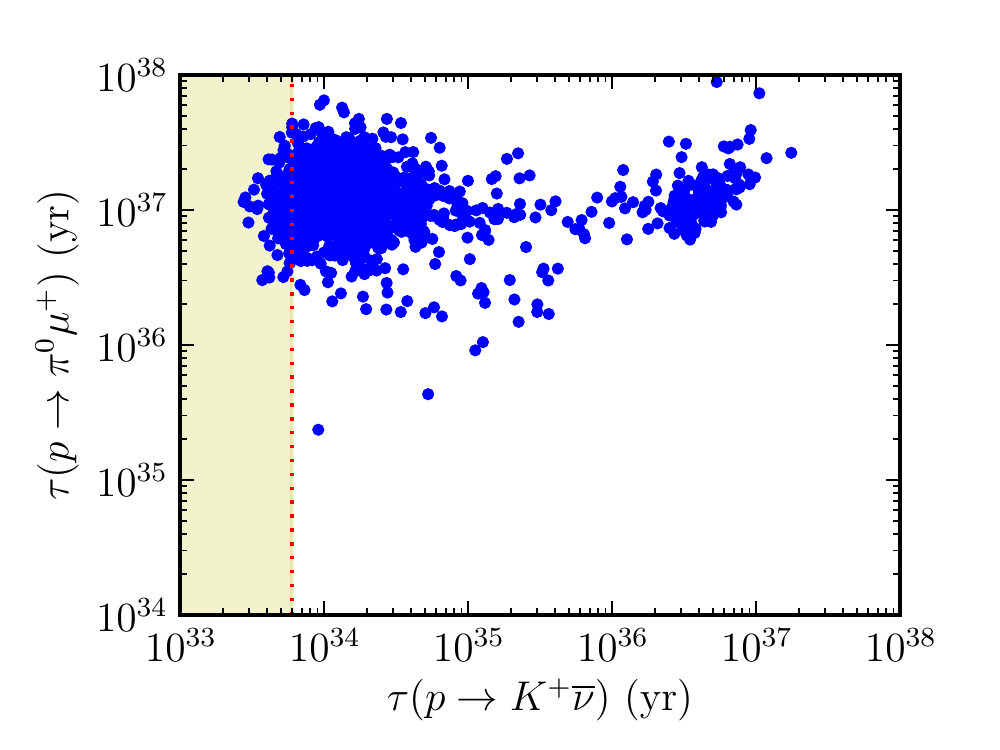}
        \includegraphics[width=7.8cm]{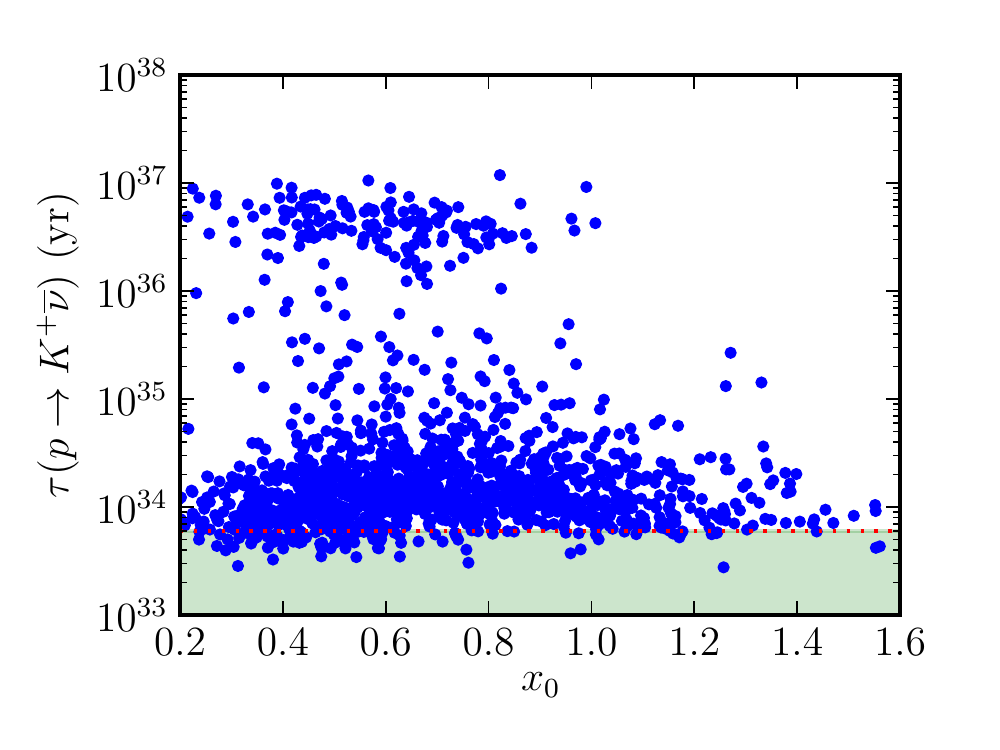}
        \caption{\footnotesize Comparisons of $K^+ \bar\nu$ partial
        lifetime to those of other dominant modes in the model, and
        that lifetime as a function of the {\bf 10} mass parameter
        $x_0$, for the type-I case. Note the unsurprising preference
        for smaller $x_0$.}
	\label{fig:KnuplotsI}
\end{figure}
\begin{figure}[t]
	\includegraphics[width=7.8cm]{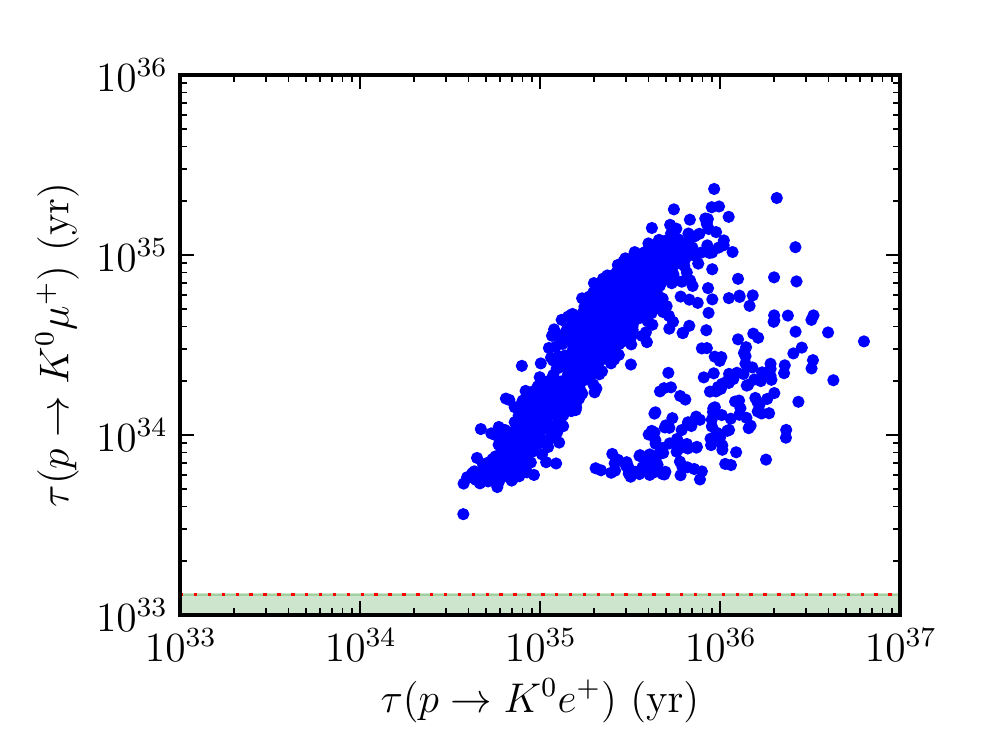} 
	\includegraphics[width=7.8cm]{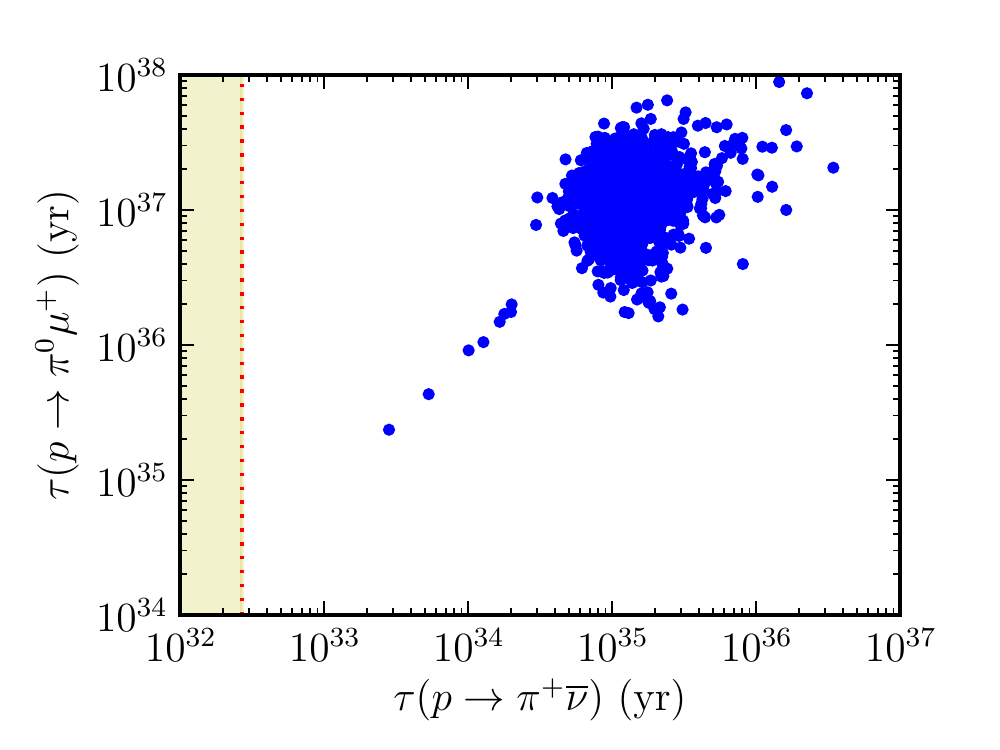}
        \caption{\footnotesize Comparisons of partial lifetimes among
        highly-correlated sub-dominant modes in the model for the
        type-I case.}
	\label{fig:otherplotsI}
\end{figure}

Due to the more favorable circumstances, I was able to locate an
allowed region of parameter space for type-I simply by running a
large number of trials with the type-II parameter seeds. I repeated
the process of expanding the range of $x_0$ by again choosing five
seeds that gave every mode sufficient and $\tau(K^+ \bar\nu)$ roughly
twice the experimental bound, and I again used those seeds to create
scatter plots for a large number of trials. Figure \ref{fig:KnuplotsI}
gives the relationships between the $K^+ \bar\nu$ mode and other
representative modes and the distribution of $\tau(K^+ \bar\nu)$ as a
function of $x_0$, and Figure \ref{fig:otherplotsI} shows the
relationships between other more closely related modes. Note the
bifurcation of the solution set in each plot; I have not yet been able
to discover the cause of this behavior.

Again I performed scans to determine a statistical upper bound for the
value of $\tau(K^+ \bar\nu)$ in the model. I consistently found that
$\tau > 10^{37}$\,years was rare and did not see a value higher than
$\sim 3\!\times\! 10^{37}$\,yr. Given those findings, I suspect that
the {\it de facto} upper limit on $\tau(K^+ \bar\nu)$ for the type-II
case is slightly lower than $10^{38}$\,years for the type-I seesaw
case.  Such a value is clearly out of reach of Hyper-K and other
imminent experiments. Note that as values for the neutral Kaon and
pion lifetimes often exceeded $10^{38}$\,years in my findings
involving $K^+ \bar\nu$ minimization, the upper limits for those modes
are surely sub-dominant to gauge exchange as well as out of reach of
experiments and so not of interest.

\section{Conclusion} \label{conclusion} In this work I have presented
a full analysis of the nature of proton decay in an $SO(10)$ model
that has {\bf 10}, $\overline{\bf{126}}$, and {\bf 120} Yukawa
couplings with restricted textures intended to naturally give
favorable results for proton lifetime as well as a realistic fermion
sector. The model is capable of supporting either type-I or type-II
dominance in the neutrino mass matrix, and I have analyzed both types
throughout. Using, numerical minimization of chi-squares, I was able
to obtain successful fits for all fermion sector parameters, including
the $\theta_{13}$ reactor mixing angle, and for both seesaw types.
Using the Yukawa couplings fixed by those fermion sector fits as
input, I then searched the parameter space of the heavy triplet Higgs
sector mixing for areas yielding adequate partial lifetimes, again
using numerical minimization to optimize results. For the case with
type-II seesaw, I found that lifetime limits for five of the six decay
modes of interest are satisfied for nearly arbitrary values of the
triplet mixing parameters, with an especially mild ${\cal O}(10^{-1})$
cancellation required in order to satisfy the limit for the $K^+
\bar\nu$ mode.  Additionally, I deduced that partial lifetime values
of $\tau(K^+ \bar\nu) \gsim 10^{36}$\,years are vanishingly unlikely
in the model, implying the value can be taken as a {\it de facto}
lifetime for the mode, which makes the model ultimately testable. For
the case with type-I seesaw, I found that limits for {\it all six}
decay modes of interest are satisfied for values of the triplet mixing
parameters that do not result in substantial enhancement, with limits
for modes other than $K^+ \bar\nu$ satisfied for nearly arbitrary
parameter values; furthermore, I deduced a statistical maximum
lifetime for $K^+ \bar\nu$ of just under $10^{38}$\,years. Given these
results, I conclude that the well-motivated Yukawa texture ansatz
proposed by Dutta, Mimura, and Mohapatra is a phenomenological
success, capable of suppressing proton decay without the usual need
for cancellation and without compromising any aspect of the
corresponding fermion mass spectrum. \newpage

\paragraph*{Acknowledgements.} This work under Rabindra Mohapatra was
supported by the University of Maryland, College Park Department of
Physics and by National Science Foundation grant number PHY-1315155. I
would like to thank R. Mohapatra for extensive discussion and guidance
and Y. Mimura and B. Dutta for helpful correspondence. I would also
like to thank M. Richman for assistance with numerical tools and
programming.

\newpage
\appendix

\begin{center}
\section{\\ Feynman Diagrams for All $d=6$ Operators Contributing to
Proton Decay} \label{fds}
\end{center}

\vspace{-4mm}
\paragraph*{\textbf{Channels for} $\boldsymbol{p \rightarrow \pi^+
\bar\nu}$.} {\footnotesize $i,l = 1,2,3$; $\tilde\phi_{\cal T}$ is the
Higgsino component of a heavy color-triplet Higgs superfield, $\phi =
{\rm H},\bar\Delta,\Sigma$}

\vspace{6mm}
\begin{figure}[h]
\begin{center}
  \includegraphics{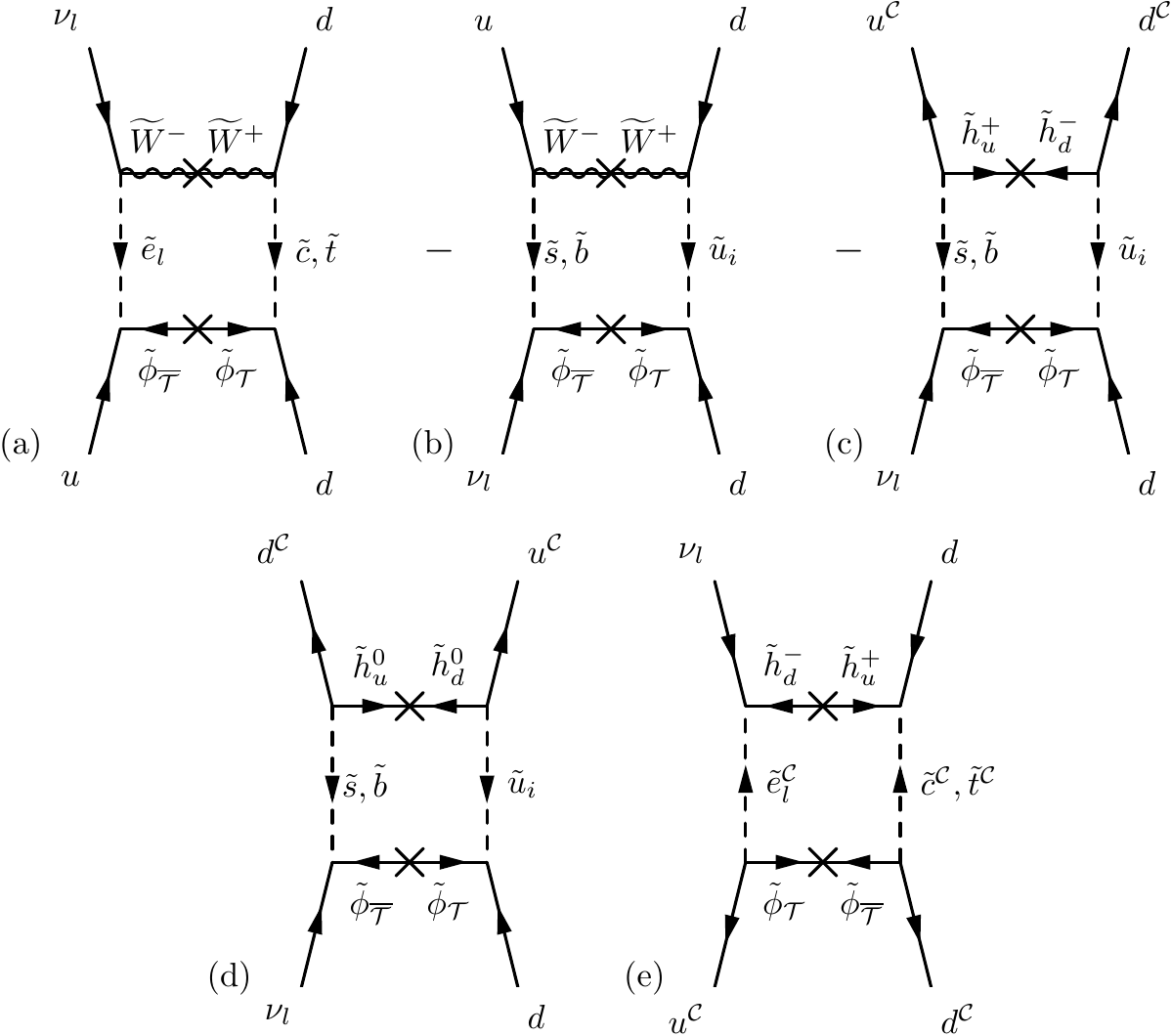}
\end{center}
\end{figure}

\newpage
\paragraph*{\textbf{Channels for} $\boldsymbol{p \rightarrow
\pi^0 \ell^+}$.} {\footnotesize $j = 1,2,3$;\, $l = 1,2$
($\leftrightarrow \ell = e,\mu$), or for diagrams including $l'$,
instead $l = 1,2,3$ and $l' = 1,2$}

\vspace{6mm}
\begin{figure}[h]
\begin{center}
  \includegraphics{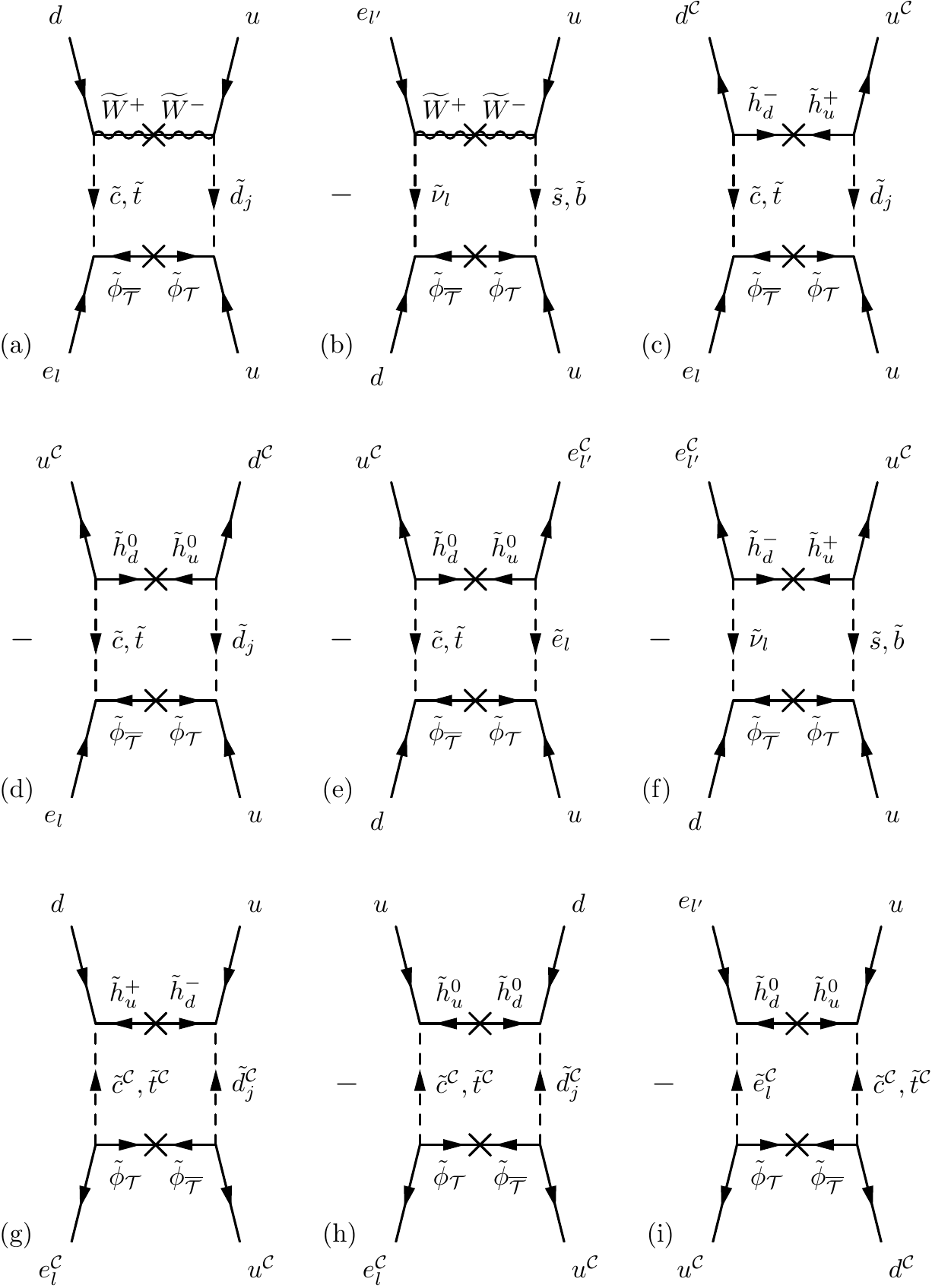}
  \vspace{-8mm}
\end{center}
\end{figure}

\newpage
\paragraph*{\textbf{Channels for} $\boldsymbol{p \rightarrow K^+
\bar\nu}$.} {\footnotesize $i,l = 1,2,3$; parentheses indicate coupled
choices; the absence of a diagrams containing $\tilde{u} d u
\tilde{e}$ dressed by $\tilde{h}^\pm$ and $u \tilde{d} d \tilde{\nu}$
dressed by $\tilde{h}^0$ is due to resulting external $\nu^{\cal C}$}

\vspace{6mm}
\begin{figure}[h]
\begin{center}
  \includegraphics{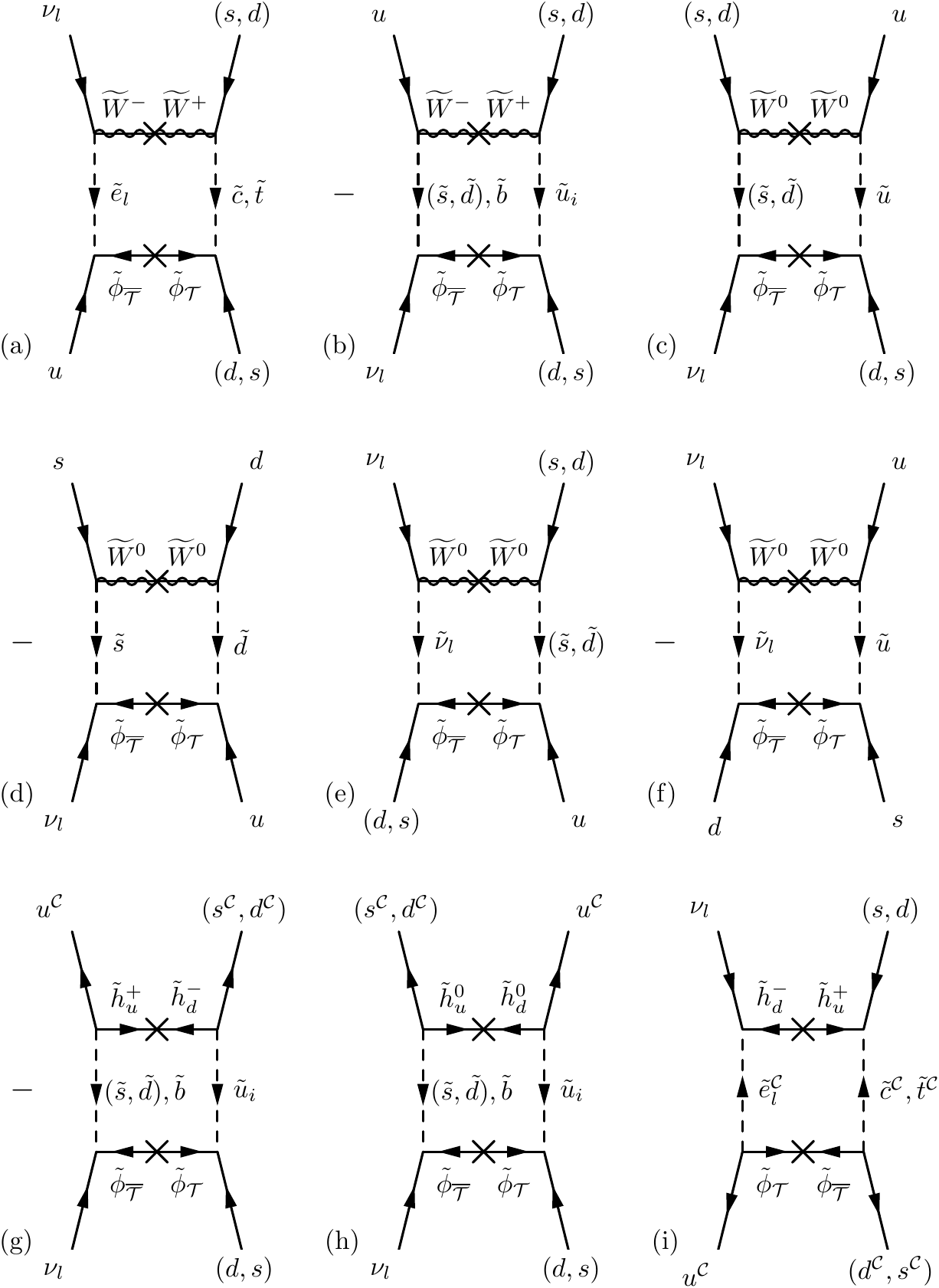}
  \vspace{-8mm}
\end{center}
\end{figure}

\newpage
\paragraph*{\textbf{Channels for} $\boldsymbol{p \rightarrow K^0
\ell^+}$.} {\footnotesize $j = 1,2,3$;\, $l = 1,2$ ($\leftrightarrow
\ell = e,\mu$), or for diagrams including $l'$, instead $l = 1,2,3$
and $l' = 1,2$}

\vspace{6mm}
\begin{figure}[h]
\begin{center}
  \includegraphics{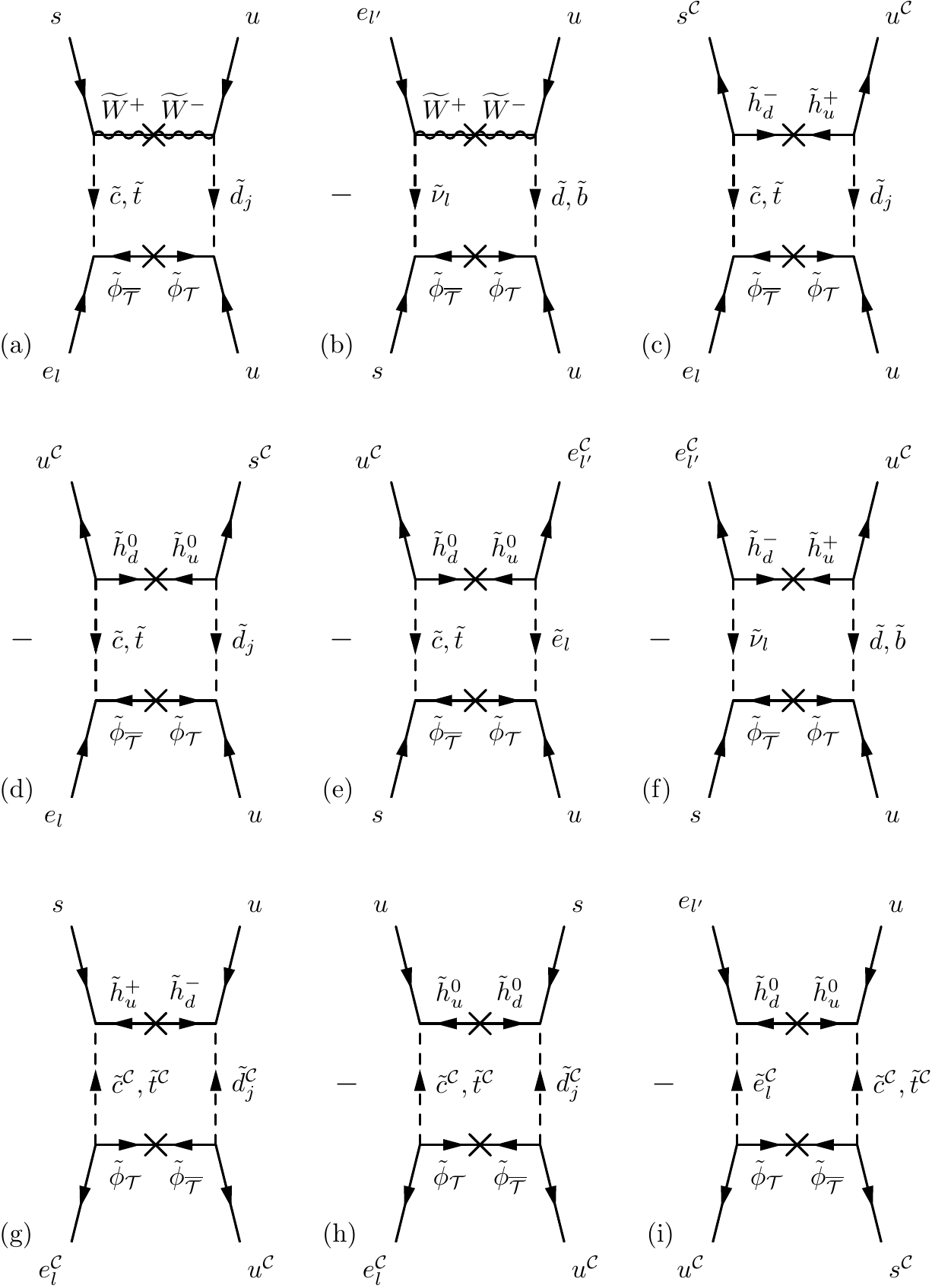}
  \vspace{-8mm}
\end{center}
\end{figure}

%
%


\newpage
\begin{thebibliography}{99}

\bibitem{minkowski} H. Fritzsch, P. Minkowski. Ann. Phys. {\bf 93}
  (1975) 93; H. Georgi, in {\it Particles and Fields} (ed. C. E.
  Carlson), A.I.P. (1975).

\bibitem{rabi-aulakh} C. S. Aulakh, R. N. Mohapatra. Phys.Rev. {\bf
  D28} (1983) 217

\bibitem{kuo} T.E. Clark, Tzee-Ke Kuo, N. Nakagawa. Phys.Lett. {\bf
  B115} (1982) 26

\bibitem{rabi-babu} K. S. Babu, R. N. Mohapatra. Phys. Rev. Lett.
  {\bf 70} (1993) 2845, [hep-ph/9209215].

\bibitem{babu-mace} K. S. Babu, C. Macesanu. Phys. Rev. {\bf D72},
  115003 (2005) [hep-ph/0505200].
  
\bibitem{10-126a} G. Lazarides, Q. Shafi, C. Wetterich. Nucl. Phys.
  {\bf B181} (1981) 287; J. Schechter, J. W. F. Valle. Phys. Rev. {\bf
  D22} (1980) 2227; R. N. Mohapatra, G. Senjanovic. Phys. Rev. {\bf
  D23} (1981) 165.

\bibitem{bajc} B. Bajc, G. Senjanovic, F. Vissani. hep-ph/0110310;
  Phys. Rev. Lett. {\bf 90}, 051802 (2003) [hep-ph/0210207].

\bibitem{goh-ng} H. S. Goh, R. N. Mohapatra, S. P. Ng. Phys. Lett.
  {\bf B570} (2003) 215, [hep-ph/0303055]; Phys. Rev. {\bf D68},
  115008 (2003) [hep-ph/0308197].

\bibitem{10-126b} S. Bertolini, M. Frigerio, M. Malinsky.  Phys. Rev.
  {\bf D70}, 095002 (2004) [hep-ph/0406117]; S. Bertolini, T.
  Schwetz, M. Malinsky. Phys. Rev. {\bf D73}, 115012 (2006)
  [hep-ph/0605006]; S. Bertolini, M. Malinsky, Phys. Rev. {\bf D72},
  055021 (2005) [hep-ph/0504241]; A. S. Joshipura, K . M. Patel.
  arXiv:1105.5943 [hep-ph].

\bibitem{fukuyama} K. Matsuda, Y. Koide, T. Fukuyama. Phys. Rev. {\bf
  D64}, 053015 (2001), [arXiv:hep-ph/0010026]; T. Fukuyama, N. Okada.
  JHEP {\bf 11} (2002) 011, [arXiv:hep-ph/0205066]; T. Fukuyama, A.
  Ilakovac, T. Kikuchi, S. Meljanac, N. Okada. JHEP {\bf 09} (2004)
  052, [arXiv:hep-ph/0406068]; T. Fukuyama, A. Ilakovac, T. Kikuchi,
  S. Meljanac, N. Okada. Eur. Phys. J. {\bf C42} (2005) 191,
  [arXiv:hep-ph/0401213].

\bibitem{seesaw} P. Minkowski. Phys. Lett. {\bf B67}, 421 (1977);
  T. Yanagida in \emph{Workshop on Unified Theories, KEK Report
  79-18} (1979) 95; M. Gell-Mann, P. Ramond, R. Slansky.
  \emph{Supergravity}, 315, Amsterdam: North Holland (1979); S.
  L. Glashow. \emph{1979 Cargese Summer Institute on Quarks and
  Leptons}, 687, New York: Plenum (1980); R. N. Mohapatra, G.
  Senjanovic. Phys. Rev. Lett. {\bf 44} (1980) 912.

\bibitem{dayabay} F. P. An {\it et al.}  [DAYA-BAY Collaboration],
  Phys. Rev. Lett. {\bf 108}, 171803 (2012) [arXiv:1203.1669
  [hep-ex]];  J. K. Ahn {\it et al.} [RENO Collaboration], Phys. Rev.
  Lett. {\bf 108}, 191802 (2012) [arXiv:1204.0626 [hep-ex]].

\bibitem{tribim} P. F. Harrison, D. H. Perkins, W. G. Scott.
  Phys.Lett. {\bf B458} (1999) 79, [hep-ph/9904297].

\bibitem{altarelli} G. Altarelli, G. Blankenburg, JHEP {\bf 1103}
  (2011) 133, [arXiv:1012.2697 [hep-ph]].

\bibitem{dms} P. S. Bhupal Dev, R. N. Mohapatra, M. Severson. Phys.
  Rev. {\bf D84}, 053005 (2011), [arXiv:1107.2378 [hep-ph]].

\bibitem{ddms} P. S. Bhupal Dev, B. Dutta, R. N. Mohapatra, M.
  Severson. Phys. Rev. {\bf D86}, 035002 (2012) [arXiv:1202.4012
  [hep-ph]].

\bibitem{superk} J. Gustafson {\it et al.}  [Super-K Collaboration],
  Phys. Rev. {\bf D91}, 072009 (2015), [arXiv:1504.01041 [hep-ex]]. 

\bibitem{dmm1302} B. Dutta, Y. Mimura, R. N. Mohapatra. Phys. Rev.
  {\bf D87}, 075008 (2013), [arXiv:1302.2574 [hep-ph]].

\bibitem{olddmm} B. Dutta, Y. Mimura, R. N. Mohapatra. Phys. Rev. 
  {\bf D72}, 075009 (2005), [arXiv:hep-ph/0507319].

\bibitem{dmmflavor}  B. Dutta, Y. Mimura, R. N. Mohapatra. JHEP {\bf
  1005} (2010) 034, [arXiv:0911.2242 [hep-ph]].

\bibitem{aulgarg} C. S. Aulakh, S. K. Garg. Nucl. Phys. {\bf B857}
  (2012) 101, [arXiv:0807.0917 [hep-ph]].

\bibitem{belyaev}  V. M. Belyaev, M. I. Vysotsky. Phys. Lett. {\bf
  B127} (1983) 215.

\bibitem{goh} H. S. Goh, R.N. Mohapatra, S. Nasri, Siew-Phang Ng.
  Phys. Lett. {\bf B587} (2004) 105, [arXiv:hep-ph/0311330].

\bibitem{pdg} K.A. Olive {\it et al.} (Particle Data Group), Chin.
  Phys. C, {\bf 38}, 090001 (2014). 

\bibitem{gavela-king} M. B. Gavela {\it et al.} Nucl. Phys. {\bf
  B312, 2} (1989) 269.

\bibitem{fukugita} S. Aoki {\it et al.} Phys. Rev. {\bf D62}, 014506
  (2000), [arXiv:hep-lat/9911026].

\bibitem{claudson} M. Claudson, M. B. Wise, L. J. Hall. Nucl. Phys.
  {\bf B195} (1982) 297.

\bibitem{donoghue} J. F. Donoghue, E. Golowich. Phys. Rev. {\bf D26}
  (1982) 3092.

\bibitem{hisano} M. Matsumoto, J. Arafune, H. Tanaka, K. Shiraishi.
  Phys. Rev. {\bf D46} (1992) 3966; J. Hisano, H. Murayama, T.
  Yanagida. Nucl. Phys. {\bf B402} (1993) 46.

\bibitem{minuit} F. James and M. Roos, Computer Physics
  Communications. {\bf 10} (1975) 343;
  http://seal.web.cern.ch/seal/MathLibs/5 10/Minuit2/html/

\bibitem{python} G. van Rossum and F.L. Drake (eds). Python Reference
  Manual, Virginia: Python Labs (2001); http://www.python.org

\bibitem{das} C.R. Das, M.K. Parida. Eur. Phys. Journal {\bf C20} (2001)
  121 [arXiv:hep-ph/0010004].

\bibitem{bora} K. Bora. {\it Horizon, A Journal of Physics}, {\bf 2}
  (2013) ISSN 2250-0871, [arXiv:1206.5909 [hep-ph]].

\bibitem{poko} T. Blazek, S. Raby, S. Pokorski. Phys. Rev. {\bf D52},
  4151 (1995) [hep-ph/9504364].

\bibitem{babuexp} K. S. Babu {\it et al.} Report of the Community
  Summer Study ({\it Snowmass 2013}), Intensity Frontier -- Baryon Number
  Violation Group, [arXiv:1311.5285 [hep-ph]].

\bibitem{hyperk} K. Abe {\it et al.}  [Hyper-K Collaboration],
  arXiv:1109.3262 [hep-ex].

\bibitem{dune} C. Adams {\it et al.}  [LBNF/DUNE Collaboration],
  arXiv:1307.7335 [hep-ex].

\end{thebibliography}
\end{document}